\documentclass[aip,amsmath,amssymb, nofootinbib,
reprint,%
]{revtex4-1}

\usepackage{graphicx}% Include figure files
\usepackage{dcolumn}% Align table columns on decimal point
\usepackage{bm}% bold math
\usepackage[utf8]{inputenc}
\usepackage[T1]{fontenc}
\usepackage{mathptmx}
\usepackage{etoolbox}
\usepackage{microtype}
\usepackage{soul}
\usepackage{xcolor}
\usepackage[dvipsnames]{xcolor}
\usepackage{makecell}
\usepackage{siunitx}
\usepackage{float}
\usepackage{appendix}
\usepackage{hyperref}
\hypersetup{
    colorlinks,
    linkcolor={black},
    citecolor={blue},
    urlcolor={black}
}

\newcommand{\qi}{Q_\mathrm{i}} % Internal quality factor inside math mode
\newcommand{\qc}{Q_\mathrm{c}} % External quality factor inside math mode
\newcommand{\kint}{\kappa_\mathrm{i}} % Internal loss rate
\newcommand{\kc}{\kappa_\mathrm{c}} % Feedline-resonator coupling rate
\newcommand{\kL}{\kappa_\mathrm{L}} % 
\newcommand{\fd}{f_\mathrm{d}} % feedline drive frequency
\newcommand{\omegad}{\omega_\mathrm{d}} % feedline drive frequency
\newcommand{\Pd}{P_\mathrm{d}} % feedline drive power
\newcommand{\fr}{f_\mathrm{r}} % Resonance frequency
\newcommand{\tr}{S_{21}} % Transmission
\newcommand{\Zr}{Z_\mathrm{r}} % Transmission
\newcommand{\nph}{\langle N_\mathrm{ph} \rangle}
\newcommand{\Bpar}{B_\parallel}
\DeclareSIUnit{\dbm}{dBm}
\usepackage{comment}%
\makeatletter
\def\@email#1#2{%
 \endgroup
 \patchcmd{\titleblock@produce}
  {\frontmatter@RRAPformat}
  {\frontmatter@RRAPformat{\produce@RRAP{*#1\href{mailto:#2}{#2}}}\frontmatter@RRAPformat}
  {}{}
}%
\makeatother
\begin{document}

\preprint{AIP/123-QED}

\title
{Low-loss frequency-tunable Josephson junction array cavities on Ge/SiGe heterostructures with a tapered etching approach}

\author{Franco De Palma$^\dagger$}
\affiliation{Hybrid Quantum Circuits Laboratory, Institute of Physics, École Polytechnique Fédérale de Lausanne (EPFL), Lausanne 1015, Switzerland}
\affiliation{Center for Quantum Science and Engineering, École Polytéchnique Fédérale de Lausanne (EPFL), Lausanne 1015, Switzerland}

\author{Elena Acinapura$^\dagger$}%
\affiliation{Hybrid Quantum Circuits Laboratory, Institute of Physics, École Polytechnique Fédérale de Lausanne (EPFL), Lausanne 1015, Switzerland}
\affiliation{Center for Quantum Science and Engineering, École Polytéchnique Fédérale de Lausanne (EPFL), Lausanne 1015, Switzerland}

\author{Wonjin Jang}%
\affiliation{Hybrid Quantum Circuits Laboratory, Institute of Physics, École Polytechnique Fédérale de Lausanne (EPFL), Lausanne 1015, Switzerland}
\affiliation{Center for Quantum Science and Engineering, École Polytéchnique Fédérale de Lausanne (EPFL), Lausanne 1015, Switzerland}

\author{Fabian Oppliger}%
\affiliation{Hybrid Quantum Circuits Laboratory, Institute of Physics, École Polytechnique Fédérale de Lausanne (EPFL), Lausanne 1015, Switzerland}
\affiliation{Center for Quantum Science and Engineering, École Polytéchnique Fédérale de Lausanne (EPFL), Lausanne 1015, Switzerland}

\author{Radha Krishnan}%
\affiliation{Hybrid Quantum Circuits Laboratory, Institute of Physics, École Polytechnique Fédérale de Lausanne (EPFL), Lausanne 1015, Switzerland}
\affiliation{Center for Quantum Science and Engineering, École Polytéchnique Fédérale de Lausanne (EPFL), Lausanne 1015, Switzerland}

\author{Arianna Nigro}
\affiliation{Physics Department, University of Basel, Klingelbergstrasse 82, Basel CH-4056, Switzerland}
\author{Ilaria Zardo}
\affiliation{Physics Department, University of Basel, Klingelbergstrasse 82, Basel CH-4056, Switzerland}
\affiliation{Swiss Nanoscience Institute, Klingelbergstrasse 82, Basel CH-4056, Switzerland}

\author{Pasquale Scarlino$^*$}
\email{pasquale.scarlino@epfl.ch}
\affiliation{Hybrid Quantum Circuits Laboratory, Institute of Physics, École Polytechnique Fédérale de Lausanne (EPFL), Lausanne 1015, Switzerland}
\affiliation{Center for Quantum Science and Engineering, École Polytéchnique Fédérale de Lausanne (EPFL), Lausanne 1015, Switzerland}

\def\thefootnote{$\dagger$}\footnotetext{These authors contributed equally to this work.}\def\thefootnote{\arabic{footnote}}

\date{\today}% It is always \today, today,
% ABSTRACT
%%%%%%%%%%%%%%%%%%%%%%%%%%%%%%%%%%%%%%%%%%%%%%%%%%%%%%%%%%%%%%%%%%%%%%%%%%%%%%%%%%%%%%%%%%%%%%%%%%%
\begin{abstract}
Ge/SiGe heterostructures represent a promising platform for hosting various quantum devices such as hole spin qubits and Andreev spin qubits. 
However, the compatibility of such heterostructures with high-quality-factor microwave superconducting cavities remains a challenge due to defects in the material stack.
%In this context, circuit quantum electrodynamics has emerged as an important tool to improve and scale up semiconductor quantum devices: the coupling of qubits with microwave photons stored in superconducting cavities enables numerous applications, including long-range interaction between distant qubits and fast and high-fidelity qubit readout.
%While such superconducting cavity-semiconductor hybrid architectures provide routes towards scalable quantum devices, defects present in Ge/SiGe planar heterostructures represent a significant loss channel for photons in superconducting cavities, which degrades the quality factor of the cavity mode and can ultimately limit the performances of quantum devices.
In this work, we present an approach to enhance the coherence of cavity modes on a reverse-graded Ge/SiGe heterostructure, which consists of etching the full $\sim \SI{1.6}{\micro\meter}$-thick Ge/SiGe stack down to its starting high-resistivity Si substrate, in order to pattern superconducting cavities directly on it. We engineer the mesa step to be tapered, so that it can be easily climbed by the superconducting cavities to reach the quantum devices potentially hosted in the Ge quantum well.
Using this approach, we observe internal quality factors of $\qi \approx 10\,000-20\,000$ for high-impedance frequency-tunable Josephson junction array resonators, limited by the junctions' fabrication, and $\qi \approx 100\,000$ for \SI{50}{\ohm} coplanar waveguide Nb lift-off resonators.
These $\qi$ are preserved despite the overlap with the mesa structure in the climbing region, and are comparable to the ones obtained for identical resonators fabricated on a high-resistivity Si wafer reference. Thereby, this work paves a practical path toward superconductor-semiconductor hybrid devices, immediately applicable to emerging technologies on planar Ge. 
\end{abstract}

\maketitle

\begin{figure*}[!ht]
    \centering
    \includegraphics{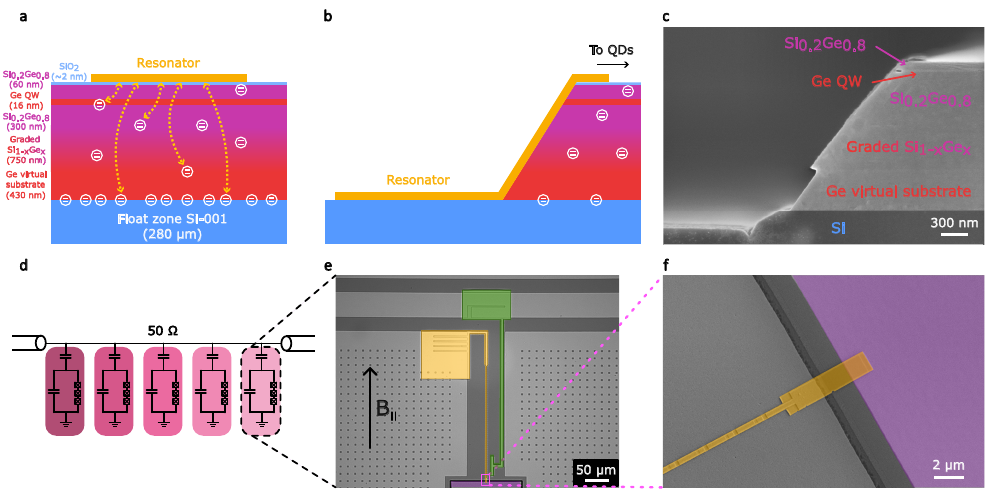}
    \caption{
    \textbf{\textsf{(a)}} Schematic of a Ge/SiGe heterostructure with a resonator defined on top of the structure. Defects, each denoted by two white parallel lines in a circle, reside in various layers and interfaces of the heterostructure, and couple to the resonator (in yellow). 
    \textbf{\textsf{(b)}} Schematic of a Ge/SiGe heterostructure etched down to the Si substrate with a tapered mesa step. The resonator is patterned mainly on the Si substrate and its termination climbs over the mesa to be connected to potential structures defined there. 
    \textbf{\textsf{(c)}} Scanning electron microscope (SEM) image of a Ge/SiGe heterostructure after the tapered etching process. The etching leaves an approximately $\SI{45}{\degree}$ angled mesa step. An approximately $\SI{100}{\nano\meter}$-deep trench is visible in the Si substrate, due to an increased ion scattering rate at the mask sidewall during the etching step (see Appendix \ref{supp:fab_JJ_arrays}). As is visible from the overetched region at the Ge virtual substrate, Ge is etched at a slightly faster rate than SiGe.
    \textbf{\textsf{(d)}} Schematic of five notch-type Josephson junction (JJ) array resonators coupled to a $\SI{50}{\ohm}$ photon feedline. 
    \textbf{\textsf{(e)}} Optical micrograph of a JJ array resonator coupled to the $\SI{50}{\ohm}$ feedline. The JJ array resonator is false-colored in yellow. One end of the resonator consists of a large pad shunting it to ground to form a quarter-wave resonator, while the other end is climbing the mesa (purple shaded structure). A hook (colored in green) extends from the feedline and defines the capacitive coupling between the feedline and the resonator.
    \textbf{\textsf{(f)}} SEM image of the JJ array resonator at the voltage antinode (yellow color). The tapered etching, in combination with the angled evaporation of the junctions, allows to successfully climb the mesa (purple color) without discontinuity of the metal.
    }
    \label{fig:devices}
\end{figure*}

%%%%%%%%%%%%%%%%%%%%%%%%%%%%%%%%%%%%%%%%%%%%%%%%%%%%%%%%%%%%%%%%%%%%%%%%%%%%
% INTRODUCTION
%%%%%%%%%%%%%%%%%%%%%%%%%%%%%%%%%%%%%%%%%%%%%%%%%%%%%%%%%%%%%%%%%%%%%%%%%%%%
Planar Ge/SiGe heterostructures have recently emerged as a prominent platform for scalable quantum computation and simulation \cite{hendrickx_fast_2020, hendrickx_single-hole_2020, hendrickx_four-qubit_2021, borsoi_shared_2024, zhang_universal_2025, wang_operating_2024}. 
2-dimensional hole gases (2DHG) in Ge quantum wells (QW) can be engineered to encompass mobilities exceeding $10^6$ \si{\centi\meter^2\per \volt \second} \cite{lodari_lightly_2022}, a small effective mass ($m_\mathrm{HH} \approx 0.05 m_e$) \cite{sammak_shallow_2019}, and low isotope populations with non-zero nuclear spins \cite{itoh_isotope_2014, moutanabbir_nuclear_2024}, all of which can be beneficial for quantum dot (QD) hole spin qubits in Ge \cite{hendrickx_fast_2020, hendrickx_four-qubit_2021, hendrickx_single-hole_2020}. 
Significantly, holes in Ge exhibit novel features such as a large spin-orbit interaction \cite{terrazos_theory_2021, hendrickx_single-hole_2020}, an anisotropic gyromagnetic ratio tensor ($g$-tensor) \cite{hendrickx_sweet-spot_2024}, and a small contact hyperfine coupling to surrounding nuclear spins \cite{fischer_spin_2008, scappucci_germanium_2021}.
These properties are also tunable via gate voltages \cite{hendrickx_sweet-spot_2024}, allowing all-electrical manipulation of the hole spin states \cite{hendrickx_fast_2020, hendrickx_single-hole_2020}, and in-situ optimization of the spin coherence \cite{hendrickx_fast_2020, hendrickx_sweet-spot_2024}. 
Additionally, superconducting germanosilicide contacts in planar Ge heterostructures \cite{tosato_hard_2023} provide novel paths to study Andreev bound states (ABSs) \cite{hinderling_direct_2024, pita-vidal_direct_2023, pita-vidal_strong_2024} and minimal Kitaev chains \cite{dvir_realization_2023} within planar Ge heterostructures, also showcasing the significance of planar Ge for various quantum device applications. 
%These advantages have recently led to remarkable progresses, including high-fidelity QD qubit operations  and scalable 2-dimensional QD array architecture , showcasing the feasibility of the Ge-based qubit platforms. 

Superconducting cavities have been used to coherently interface microwave photons with various quantum systems \cite{blais_circuit_2021}.
In this regard, embedding semiconductor quantum states into superconducting cavities unlocks possibilities for novel quantum architectures, facilitating, for instance, high-fidelity spin readout \cite{zheng_rapid_2019}, coherent interconnects between remote spins\cite{harvey-collard_coherent_2022, dijkema_cavity-mediated_2025}, microwave optics \cite{oppliger_high-efficiency_2025, liu_semiconductor_2015}, and readout and manipulation of Andreev bound states \cite{bargerbos_spectroscopy_2023, pita-vidal_direct_2023, wesdorp_microwave_2024}.
In such hybrid circuit quantum electrodynamics (cQED) approach \cite{burkard_superconductorsemiconductor_2020}, achieving large hole-photon coupling strengths and low photon loss rates (or large internal quality factors $\qi$) of the resonators are needed for realizing coherent hole-photon interfaces \cite{de_palma_strong_2024, yu_strong_2023, janik_strong_2025}. 
In this context, recent studies have demonstrated strong hole charge-photon coupling in Ge/SiGe hetrostructures, exploiting a high-impedance superconducting resonator realized with a SQUID array \cite{de_palma_strong_2024} and a high-kinetic granular aluminum (grAl) nanowire \cite{janik_strong_2025}.
However, a substantial amount of defects present in Ge/SiGe heterostructures greatly limits the internal quality factor of the cavity modes in this platform to $\qi \approx 3\,000-5\,000$ for forward-graded heterostructures \cite{valentini_parity-conserving_2024, sagi_gate_2024, janik_strong_2025} and $\qi \approx 1\,000$ for reversed-graded ones \cite{nigro_demonstration_2024} in the low-photon regime, relevant in some cQED applications \cite{blais_circuit_2021, burkard_superconductorsemiconductor_2020}.

In this work, we present an approach to significantly reduce the internal losses of superconducting cavities on planar Ge. It consists of a tapered etching of the full heterostructure down to the intrinsic Si substrate, leaving only a small mesa on planar Ge for hosting semiconductor or hybrid super-semi devices.
We fabricate superconducting high-impedance ($\Zr = \SI{3}{\kilo\ohm}$) Al/AlOx/Al Josephson junction (JJ) array resonators \cite{masluk_microwave_2012} on the etched heterostructure, which also climb over the tapered mesa step, and show that the internal quality factor of the resonators increases from $\qi \approx 3\,000-4\,000$ on the bare Ge/SiGe heterostructure, to $\qi \approx 10\,000-20\,000$ on the etched structure. These values match the ones obtained on an intrinsic Si substrate reference, suggesting that the obtained upper bound in the $\qi$ is not limited by the tapered etching process, but rather by the Josephson junctions fabrication process.
We confirm this by fabricating low-impedance ($\Zr = \SI{50}{\ohm}$) coplanar waveguide (CPW) Nb lift-off resonators on the etched Ge/SiGe heterostructure which exhibit $\qi \approx 100\,000$ at low-photon number.
%we further show that the $\qi \approx 10\,000 - 20\,000$ of the JJ array resonators on the etched structure is mainly limited by the defects residing in the AlOx , rather than those in the substrate.
Furthermore, we tune the inductance of the JJ arrays resonators with an in-plane magnetic field parallel to the JJ arrays and effectively modulate their resonance frequency (and hence the impedance) in a wide-range of $\qtyrange[range-units = single, range-phrase = -]{4}{7}{\GHz}$ ($\qtyrange[range-units = single, range-phrase = -]{5}{3}{\kilo\ohm}$), which represents a useful resource for studying hybrid systems \cite{oppliger_high-efficiency_2025}.
We also characterize a degradation of the $\qi$ down to $\approx 3\,000$ as the magnetic field approaches the junctions' critical field, which can be attributed either to an increase of the quasiparticle population near such critical field \cite{scigliuzzo_phononic_2020, frasca_nbn_2023}, or also to the creation of vortices in the proximity of the resonators \cite{nsanzineza_trapping_2014}.
This work provides a systematic path toward optimizing a hybrid cQED architecture on planar Ge, which can be highly beneficial for various quantum applications involving hole-photon interactions \cite{hinderling_direct_2024, janik_strong_2025, de_palma_strong_2024}.

\begin{figure*}[!ht]
    \centering
    \includegraphics[width=\linewidth]{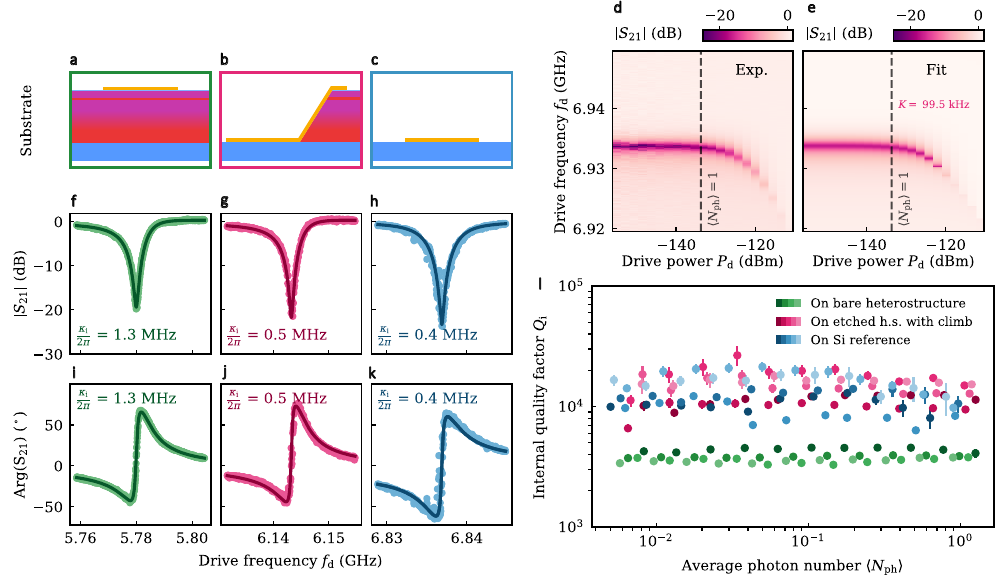}
    \caption{Spectroscopy of JJ array resonators fabricated on different substrates. 
        Schematic of the cross-section of the resonator on \textbf{\textsf{(a)}} a bare Ge/SiGe heterostructure, \textbf{\textsf{(b)}} a Ge/SiGe heterostructure etched down to Si with one resonator's end climbing the tapered mesa, and \textbf{\textsf{(c)}} an intrinsic Si substrate.
        \textbf{\textsf{(d)}} Normalized feedline transmission amplitude $|S_{21}|$ as a function of the resonator drive frequency $\fd$ and drive power $\Pd$ for a resonator on the etched heterostructure (panel \textsf{(b)}).
        \textbf{\textsf{(e)}} Simulated $|S_{21}|$ as a function of $f_\mathrm{d}$ and $P_\mathrm{d}$, with the parameters extracted from the numerical fit of \textsf{(d)} to an input-output model taking into account the Kerr-nonlinearity of the resonator (see Appendix~\ref{supp:fitting_nonlinear}).
        Dashed lines in \textsf{(d)} and \textsf{(e)} denote the power  $\Pd^\mathrm{SP} \approx \SI{-133}{\dbm}$ corresponding to an average resonator photon number $\nph \approx 1$.
        \textbf{\textsf{(f-h)}} $|\tr|$ measured as a function of $\fd$ for the respective resonator illustrated in panels \textsf{(a-c)}.
        \textbf{\textsf{(i-k)}} Phase of the normalized feedline transmission $\mathrm{Arg}(\tr)$ measured as a function of $\fd$ for the respective resonator illustrated in panels \textsf{(a-c)}.
        Solid lines in \textsf{(f-k)} are fits to an input-output model (see Appendix~\ref{supp:fitting_linear}), yielding the internal resonator loss rates $\kint$ reported in \textsf{(f-k)}.
        The power of the drive tone at the feedline $\Pd$ is $\SI{-140}{\dbm}$ for all the panels \textsf{(f-k)}.
        \textbf{\textsf{(l)}} Internal quality factor $\qi$ as a function of $\nph$ for the resonators respectively on the bare Ge/SiGe heterostructure (green data), on the Ge/SiGe heterostructure etched down to Si with one resonator end climbing the tapered mesa (pink data), and on intrinsic Si (blue data). 
        Different shades of each color denote different resonators on the same type of substrate.
        The value of $\qi$ for the different resonators are extracted from power sweeps similar to the one presented in panel \textsf{(d)} (see Appendix~\ref{supp:power_sweeps_all}).
        }
    \label{fig:resonances}
\end{figure*}

%%%%%%%%%%%%%%%%%%%%%%%%%%%%%%%%%%%%%%%%%%%%%%%%%%%%%%%%%%%%%%%%%%%%%%%%%%%%
% DESCRIPTION OF THE PROBLEM
%%%%%%%%%%%%%%%%%%%%%%%%%%%%%%%%%%%%%%%%%%%%%%%%%%%%%%%%%%%%%%%%%%%%%%%%%%%%

Figure~\ref{fig:devices}a schematically represents the cross section of a reverse-graded Ge/SiGe heterostructure that hosts defects (white circles) throughout the material stack \cite{sammak_shallow_2019, nigro_high_2024}. 
These defects can originate from amorphous materials, such as oxides at the top interface \cite{mendes_martins_defect_2023}, impurities due to contaminants during the growth process \cite{sammak_shallow_2019, nigro_high_2024, shah_reverse_2008}, and threading dislocations due to the lattice constant mismatch at the interfaces between different materials in the stack \cite{lee_strained_2004, scappucci_germanium_2021,sammak_shallow_2019}.
These can effectively act as two-level fluctuators (TLFs), or result in buried conductive layers hosting free charge carriers introducing resistive losses for the resonators .
The microwave electric and magnetic field confined in superconducting cavities can couple to these defects, effectively opening up channels for photon dissipation \cite{martinis_decoherence_2005, de_sousa_dangling-bond_2007, burnett_evidence_2014, nigro_demonstration_2024, de_graaf_suppression_2018}. 
As a result, superconducting resonators defined on semiconductor heterostructures typically exhibit internal quality factors few orders of magnitude smaller than the ones on high-resistivity substrates such as intrinsic Si \cite{tominaga_intrinsic_2025}, which in contrast can reach up to $\qi \approx 10^6$. 

One of the main photon loss mechanisms on reverse-graded Ge/SiGe heterostructures most likely stems from residual conductive layers located at the interface between the bottom Si substrate and the Ge virtual substrate, due to the lattice mismatch \cite{nigro_demonstration_2024}. 
For this reason, etching completely the heterostructure down to the bottom intrinsic Si substrate, and defining the resonators directly on it, can effectively eliminate the majority of the loss channels.
In our work, we selectively etch the Ge and SiGe layers ($\approx\SI{1.6}{\micro\meter}$ thick) down to the bottom high-resistive Si substrate (resistivity $ > \SI{10}{\kilo\ohm\cdot\cm}$, see Appendix~\ref{supp:growth}), leaving only relatively small mesa islands ($\qtyproduct{150 x 150}{\micro\meter}$) for potentially hosting semiconductor devices in the Ge QW. 
The etch process results in a surface with an extremely low roughness of $\approx \SI{0.1}{\nano\meter}$ on the Si substrate and is engineered to provide a tapered profile on the mesa step, facilitating connections from the bottom Si to the top SiGe and further fabrication processing, as illustrated in Fig.~\ref{fig:devices}b. 
Details about the etch process are provided in Appendix~\ref{supp:etching}.
Fig.~\ref{fig:devices}c reports a scanning electron microscope (SEM) image of the cross-section of the heterostructure after the etch process. Despite the Ge virtual substrate is etched slightly faster than SiGe, the profile shows a smoothly tapered step \cite{oehrlein_selective_1991}.

%%%%%%%%%%%%%%%%%%%%%%%%%%%%%%%%%%%%%%%%%%%%%%%%%%%%%%%%%%%%%%%%%%%%%%%%%%%%
% ANTICIPATION OF RESULTS
%%%%%%%%%%%%%%%%%%%%%%%%%%%%%%%%%%%%%%%%%%%%%%%%%%%%%%%%%%%%%%%%%%%%%%%%%%%%
To investigate the microwave properties of the etched heterostructure, we fabricate high-impedance ($\Zr = \SI{3}{\kilo\ohm}$) Al/AlOx/Al JJ array resonators on top of the Si substrate in the etched region. As depicted in Fig.~\ref{fig:devices}d, we couple five notch-type JJ array resonators, each with a distinct number of JJs, to a single $\SI{50}{\ohm}$ photon feedline \cite{chen_scattering_2022, probst_efficient_2015, mcrae_materials_2020}.
The JJ arrays are defined with the standard Dolan-bridge technique\cite{frunzio_fabrication_2005}, using two angled evaporations at $\pm \ang{45}$ \cite{dolan_offset_1977} (see Appendix~\ref{supp:fab_JJ_arrays}). The number of JJs in a resonator ($50 \geq N \geq 42$) determines its total inductance, and therefore its resonance frequency $\fr$ ($\qty{6}{GHz} \leq \fr \leq \qty{7}{GHz}$). From room temperature resistance measurements of junction test structures, we extract a resistance per JJ of $R_\mathrm{JJ} \approx \SI{1.25}{\kilo\ohm}$, corresponding to an inductance of $L_\mathrm{JJ} \approx \SI{1.48}{\nano\henry}$, and a critical current of $I_\mathrm{c,JJ} \approx \SI{220}{\nano\ampere}$ \cite{ambegaokar_tunneling_1963}. Given the JJ area, we also estimate a ratio between Josephson and charging energy $E_\mathrm{J}/E_\mathrm{C}$ of about 180.
%, which ensures that the $\qi$ is not limited by phase slips in the array\cite{kuzmin_quantum_2019} (see Appendix~\ref{supp:design_JJ_arrays} for further details on the JJ array resonator design parameters).

%%%%%%%%%%%%%%%%%%%%%%%%%%%%%%%%%%%%%%%%%%%%%%%%%%%%%%%%%%%%%%%%%%%%%%%%%%%%
% DESCRIPTION OF FIGURE 1
%%%%%%%%%%%%%%%%%%%%%%%%%%%%%%%%%%%%%%%%%%%%%%%%%%%%%%%%%%%%%%%%%%%%%%%%%%%%
Figure~\ref{fig:devices}e shows an optical micrograph of one of the notch resonators, where the JJ array (false-colored in yellow) is shunted to the ground plane at one end, and capacitively coupled to the photon feedline at the other end via a hook structure (false-colored in green), resulting in a quarter-wave resonator \cite{blais_circuit_2021}.
The capacitive coupling between the hook and the resonator determines the external coupling rate $\kc$ and the external quality factor $\qc = 2\pi \fr / \kc $.
To facilitate the galvanic connection between the resonator and potential structures on top of the mesa, we terminate the narrow JJ array with a wider and longer junction, which does not significantly contribute to the total inductance.
%Furthermore, the resonator has to be galvanically connected to the structures defined on the planar Ge to allow effective hole-photon interaction in a cQED device .
In Fig.~\ref{fig:devices}f we show that the angled evaporation allows a successful climbing of the tapered mesa structure (shaded in purple) without further optimization of the fabrication steps.
We also probe the electrical conduction of twenty similar test structures climbing the mesa, which are fabricated on the same chip, and find that the climbing does not introduce any disconnection, as can be anticipated from the smooth etch profile shown in Fig.~\ref{fig:devices}c.

%%%%%%%%%%%%%%%%%%%%%%%%%%%%%%%%%%%%%%%%%%%%%%%%%%%%%%%%%%%%%%%%%%%%%%%%%%%%
% MEASUREMENTS OF JJ ARRAYS
%%%%%%%%%%%%%%%%%%%%%%%%%%%%%%%%%%%%%%%%%%%%%%%%%%%%%%%%%%%%%%%%%%%%%%%%%%%%

To systematically quantify the effect of this etching approach on the internal loss rates of JJ array resonators, we fabricate nominally identical resonators on the bare Ge/SiGe heterostructure (Fig.~\ref{fig:resonances}a), on the etched Ge/SiGe heterostructure (Fig.~\ref{fig:resonances}b) and on an intrinsic (resistivity $ \approx \SI{10}{\kilo\ohm\cdot\cm}$) Si reference substrate (Fig.~\ref{fig:resonances}c). We characterize the resonators by performing microwave measurements in a dilution refrigerator at a base temperature of \qty{10}{\milli\kelvin}. See Appendix~\ref{supp:measurement_setup} for details about the measurement setup. 

We first explore the power dependence of the resonators' spectrum.
Fig.~\ref{fig:resonances}d presents the normalized feedline transmission amplitude $|\tr|$ as a function of the estimated drive power $\Pd$ at the sample feedline (see Appendix~\ref{supp:measurement_setup}) and drive frequency $\fd$, measured around the resonance frequency of one of the five notch resonators on the etched heterostructure with the mesa climb (Fig.~\ref{fig:resonances}b). 
For the power spectroscopies of all resonators on all substrates, refer to Appendix~\ref{supp:power_sweeps_all}. 
Starting from %a power of about 
$\Pd \approx \qty{-130}{dBm}$, a decrease of the resonance frequency for increasing $\Pd$ becomes distinguishable, which is attributed to the Kerr nonlinearity of the JJs \cite{eichler_controlling_2014, anferov_millimeter-wave_2020}. 
We fit the complex spectroscopy data at each individual power $\Pd < \qty{-130}{dBm}$ to a linear input-output theory model for notch-type resonators (see Appendix~\ref{supp:fitting_linear}), from which we extract the resonance frequency $\fr$, the external coupling rate $\kc$, the internal loss rate $\kint$  and the average photon number $\nph$ at each value of $\Pd$. 
From the fit we extract that $\nph = 1$ for $\Pd \approx \qty{-133}{dBm}$.
%which is found to match with the range of $\Pd$ at which the Kerr-induced frequency bending becomes distinguishable. 
We further fit numerically the full 2D spectroscopy data (Fig.~\ref{fig:resonances}j) to an input-output model that takes into account the self-Kerr nonlinearity (see Appendix~\ref{supp:fitting_nonlinear}), as presented in Fig.~\ref{fig:resonances}e, and extract the self-Kerr coefficient $K = \qty{99.5 \pm 0.4}{\kilo\hertz}$.
This value is consistent with the expected one for $N = 44$ junctions and a single-junction charging energy $E_\mathrm{C}\approx\qty{0.6}{\giga\hertz}$ (see Appendix~\ref{supp:power_sweeps_all}).

We perform similar measurements for all the resonators on the three different substrates (see Appendix~\ref{supp:power_sweeps_all}), and report their spectrum measured at a specific value of $\nph < 1$ in Figs.~\ref{fig:resonances}f-k, for one representative resonator on each substrate type indicated in Figs.~\ref{fig:resonances}a-c.
Solid lines in Figs.~\ref{fig:resonances}f-k present the numerical fit to the linear input-output model for each spectrum.
Figure~\ref{fig:resonances}l shows the $\qi$ for all the resonators, extracted from a numerical fit to the linear input-output model at the low-photon-number regime $\nph < 1$, relevant in some cQED applications\cite{blais_circuit_2021, burkard_superconductorsemiconductor_2020} (see Appendix~\ref{supp:Qc_vs_Nph} for the extracted $\qc$ of all the resonators). 
For $\nph > 1$, the large self-Kerr nonlinearity of the JJ array resonators complicates an accurate estimation of the $\qi$, as explained in Appendix~\ref{supp:fitting_nonlinear}.
In Figure~\ref{fig:resonances}l, different shades of the same color denote the different resonators on the same substrate with a distinct number of JJs. The resonators on the same substrate exhibit consistent values of $\qi$ regardless of the number of JJs. We observe that the $\qi$ remain relatively constant over the range of $\nph$ shown in Fig.~\ref{fig:resonances}l. This is expected when $\nph \ll n_\mathrm{c}$, with $n_\mathrm{c}$ the critical photon number at which the TLF-induced losses would start becoming negligible compared to other losses\cite{burnett_analysis_2016, scigliuzzo_phononic_2020}.

Notably, Fig.~\ref{fig:resonances}l shows that the resonators on the etched heterostructure feature $\qi \approx 10\,000 - 20\,000$ (pink data), systematically outperforming those on the bare heterostructure showing $\qi \approx 3\,000-4\,000$ (green data).
This clearly demonstrates that the etching technique is effectively eliminating potential loss channels residing in the heterostructure, enhancing the lifetime of the photons in the resonators.

\begin{figure}[t]
    \centering
    \includegraphics{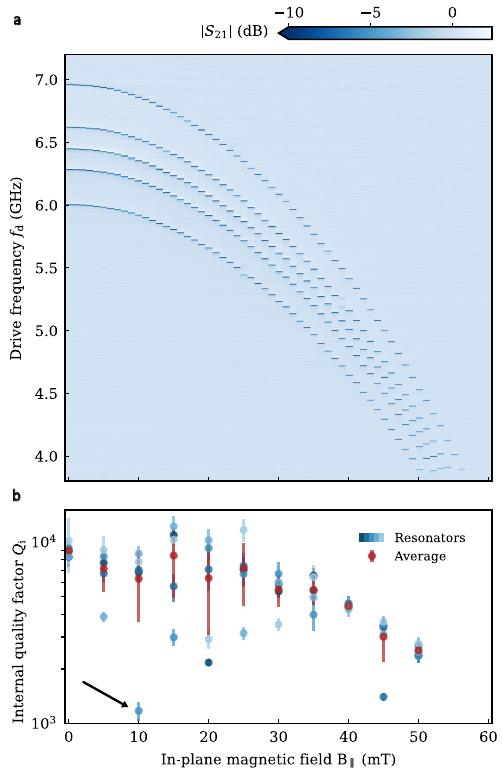}
    \caption{Magnetospectroscopy of the resonators on the intrinsic Si substrate.
    \textbf{\textsf{(a)}} Normalized feedline transmission amplitude $|S_{21}|$ measured as a function of the resonator drive frequency $\fd$ and the in-plane magnetic field $B_\parallel$ parallel to the JJ arrays. Five JJ array resonators with a different number of JJs are coupled to the same feedline, as depicted in Fig.~\ref{fig:devices}d. 
    % The red dashed lines superimposed to the experimental data show a fit of the resonance frequencies as a function of $B_\mathrm{\parallel}$ to a model taking of critical current modulation due to the field penetration into the junctions.
    \textbf{\textsf{(b)}} Internal quality factor $\qi$ of the resonators, averaged over multiple drive powers in the low-photon-number regime as a function of $B_\parallel$ (blue data). Red data points represent the $\qi$ averaged over the five resonators.
    The black arrow indicates a resonator with a reduced $\qi$ due to the coupling with a TLF (see Appendix~\ref{supp:strong_hybridization}).
    }
    \label{fig:magnetic_field}
\end{figure}

Furthermore, even though there is a finite region where the microwave cavity is inevitably overlapping with the mesa, the $\qi$ on the etched structure (pink data) match those on the intrinsic Si reference (blue data), and are hence limited by loss mechanisms associated to typical Dolan-style JJs, such as TLFs in the junctions' tunneling oxide, lift-off residues, or TLFs in the native SiO$_\mathrm{2}$ or in the Si substrate \cite{van_harlingen_decoherence_2004, constantin_microscopic_2007, zanuz_mitigating_2024}.
%Moreover, the area of the overlapping region $A_\mathrm{overlap}$ of the resonator with the mesa is kept minimal ($A_\mathrm{overlap} \approx \qtyproduct{2 x 5}{\micro\meter}$). 
The combination of this with a minimal overlapping area of the resonators with the mesa ($A_\mathrm{overlap} \approx \qtyproduct{2 x 5}{\micro\meter}$) makes it difficult to estimate the residual losses potentially introduced by the etching process itself and the overlap with the mesa.
For completeness, we conduct electromagnetic (EM) simulations to demonstrate that our tapered etching approach can also be effective for other Ge/SiGe heterostructures that initially exhibit even higher microwave losses, with the simulated $\qi$ showing significant dependence on the overlap area (See Appendix~\ref{supp:simulations_losses}).

%%%%%%%%%%%%%%%%%%%%%%%%%%%%%%%%%%%%%%%%%%%%%%%%%%%%%%%%%%%%%%%%%%%%%%%%%%%%
% MAGNETIC FIELD BEHAVIOUR
%%%%%%%%%%%%%%%%%%%%%%%%%%%%%%%%%%%%%%%%%%%%%%%%%%%%%%%%%%%%%%%%%%%%%%%%%%%%
%For cQED applications with spin-based semiconductor devices, the behaviour of superconducting resonators in an external magnetic field plays a fundamental role. 
In this work, we also characterize JJ array resonators in the presence of an in-plane magnetic field $\Bpar$ up to \qty{60}{\milli\tesla}, applied parallel to the JJ arrays' axis (see Fig.~\ref{fig:devices}e). Such magnetic field range is compatible with the operation of spin qubits in planar Ge \cite{zhang_universal_2025, jirovec_singlet-triplet_2021, hendrickx_sweet-spot_2024}.
Figure~\ref{fig:magnetic_field}a shows the spectroscopy measurement of the five notch resonators on the intrinsic Si substrate as a function of $\Bpar$, where $\fr$ of each resonator is effectively tuned in a wide range of $7-\SI{4}{GHz}$ in a span of $\Bpar < \SI{60}{\milli\tesla}$. For magnetic fields above \qty{60}{mT} the resonance frequencies are below the measurement bandwidth (see Appendix~\ref{supp:measurement_setup}).
The observed modulation of $\fr$ is due to an effective variation of $L_\mathrm{JJ}$, which can be attributed to the interplay of two different effects: (a) the reduction of the superconducting gap of Al with an increasing external magnetic field, and (b) the penetration of $\Bpar$ inside the junctions, generating a magnetic flux that modulates their critical current\cite{krause_magnetic_2022, kuzmin_tuning_2023}. An analytical model for $\fr(\Bpar)$ and a fit of the experimental data is provided in Appendix~\ref{supp:B_field_behaviour}, where we extract the junctions' in-plane critical field to be approximately $B^\parallel_\mathrm{crit} \approx \qty{66}{mT}$.

Furthermore, we characterize the $\qi$ of the resonators as a function of $\Bpar$ in steps of $\SI{5}{\milli\tesla}$ by performing resonator spectroscopy as a function of the microwave drive power $\Pd$ in the low-photon-number regime ($\nph < 1$). 
Figure~\ref{fig:magnetic_field}b shows the average $\qi$ of each resonator in the low-photon-number regime (blue data points) as a function of $\Bpar$, while the red data points represent the $\qi$ averaged over the five resonators.
Some resonators exhibit a strong reduction in $\qi$ for specific values of $\Bpar$ (see for example the data point indicated by the black arrow in Fig.~\ref{fig:magnetic_field}b), which we attribute to some TLFs in the junctions' tunneling barrier becoming resonant with the resonator mode and strongly coupling to it \cite{constantin_microscopic_2007, van_harlingen_decoherence_2004, de_palma_strong_2024}. We report such an example in Appendix~\ref{supp:strong_hybridization}. We observe an overall trend of decreasing $\qi$ with increasing $\Bpar$, which can be attributed to an increased quasiparticle population due to the superconducting gap reduction as $\Bpar$ approaches the critical field of the JJ Al leads \cite{scigliuzzo_phononic_2020, frasca_nbn_2023}, or to the creation of vortices in the proximity of the resonators \cite{nsanzineza_trapping_2014}. 
Nevertheless, the average $\qi$ stays above $5\,000$ for magnetic fields up to $\SI{40}{\milli\tesla}$. 

%%%%%%%%%%%%%%%%%%%%%%%%%%%%%%%%%%%%%%%%%%%%%%%%%%%%%%%%%%%%%%%%%%%%%%%%%%%%
% CPW RESONATORS
%%%%%%%%%%%%%%%%%%%%%%%%%%%%%%%%%%%%%%%%%%%%%%%%%%%%%%%%%%%%%%%%%%%%%%%%%%%%
To further probe the limit of the photon loss rates of resonators on the etched Ge/SiGe heterostructure, we also characterize quarter-wave $\SI{50}{\ohm}$ Nb CPW resonators. Moreover, such $\SI{50}{\ohm}$ superconducting resonators are relevant for various cQED architectures, including for instance superconducting qubits \cite{sagi_gate_2024} and Andreev spin qubits \cite{bargerbos_spectroscopy_2023, wesdorp_microwave_2024, pita-vidal_strong_2024} systems, which do not necessitate large fluctuations of the vacuum electric field in the resonator to approach the strong coupling regime.
\begin{figure}[t]
    \centering
    \includegraphics{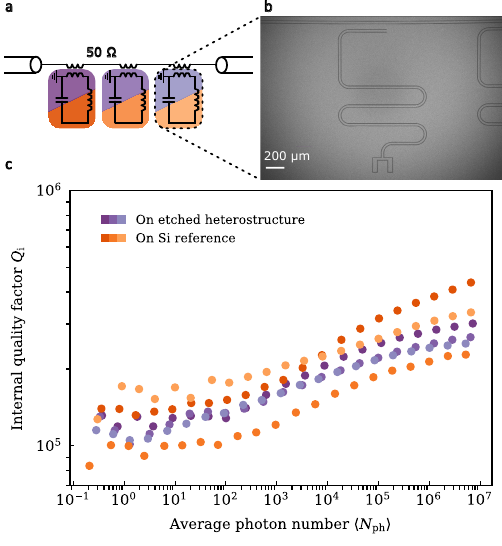}
    \caption{$\SI{50}{\ohm}$ Nb superconducting resonators on an etched Ge/SiGe heterostructure and on intrinsic Si.
    \textbf{\textsf{(a)}} Schematic of three notch-type quarter-wave resonators inductively coupled to a $\SI{50}{\ohm}$ feedline. 
    \textbf{\textsf{(b)}} Scanning electron microscope (SEM) of a $\SI{50}{\ohm}$ Nb resonator on the etched Ge/SiGe heterostructure. 
    \textbf{\textsf{(c)}} Internal quality factor $\qi$ of the $\SI{50}{\ohm}$ Nb resonators on the etched Ge/SiGe heterostructure (purple) and on intrinsic Si (orange), as a function of the average number of photons in the resonator $\langle N_\mathrm{ph} \rangle$. Different shades of each color denote different resonators on the same type of substrate.
    }
    \label{fig:cpw}
\end{figure}
As illustrated in Fig.~\ref{fig:cpw}a,b, we fabricate three lift-off \qty{50}{\nano\meter}-thick CPW Nb resonators with different lengths inductively coupled to the same $\SI{50}{\ohm}$ photon feedline in a notch configuration. The same resonators are fabricated both on the etched heterostructure and on the intrinsic Si substrate. See Appendix~\ref{supp:fab_CPW} for details about the fabrication.
%Figure~\ref{fig:cpw}b shows an SEM image of one of the resonators on the etched heterostructure, where one end of the resonator closer to the feedline on top is shunted to the ground plane to allow the inductive coupling between the feedline and the resonator. The other end of the resonator is electrically floating with respect to the ground plane potential, resulting in a quarter-wave resonator (see Appendix~\ref{supp:design_CPW}). 
Differently from the JJ array resonators, the CPW resonators on the etched heterostructure presented here do not climb any mesa island, providing an upper limit for the $\qi$ on an etched surface and with a lift-off fabrication method. 

Figure~\ref{fig:cpw}c reports $\qi$ as a function of $\nph$ for the resonators on the etched heterostructure (purple dots) and on the intrinsic Si substrate (orange dots).
To extract $\qi$ shown in Fig.~\ref{fig:cpw}c, we perform spectroscopy of each resonator as a function of $\Pd$, and numerically fit to a linear input-output model for each value of $\Pd$.
We note that the absence of non-linearity of the CPW resonators facilitates the numerical fit to the linear model up to $\nph \gg 1$, as opposed to the case of JJ array resonators (Fig.~\ref{fig:resonances}l), where the fit to the linear model was limited to the range $\nph < 1$.
In Fig.~\ref{fig:cpw}c, we observe $\qi \approx 100\,000$ for $\nph \approx 1$ for both substrates, which is about two orders of magnitude larger than the $\qi$ obtained for similar resonators defined directly on top of the Ge/SiGe heterostructure \cite{nigro_demonstration_2024}.
The presented internal quality factors increase to $\qi \approx 400\,000$ for $\nph \gg 1$, implying TLF-dominated losses, most likely due to the native SiO$_2$ and NbOx \cite{burnett_analysis_2016,  burnett_evidence_2014}. 
Here, the extracted similar values of $\qi$ on the etched heterostructure and on the intrinsic Si clearly indicate that the etching process itself does not introduce additional loss channels for CPW resonators.

%%%%%%%%%%%%%%%%%%%%%%%%%%%%%%%%%%%%%%%%%%%%%%%%%%%%%%%%%%%%%%%%%%%%%%%%%%%%
% CONCLUSIONS
%%%%%%%%%%%%%%%%%%%%%%%%%%%%%%%%%%%%%%%%%%%%%%%%%%%%%%%%%%%%%%%%%%%%%%%%%%%%
In conclusion, in this work we perform a tapered etching of a Ge/SiGe heterostructure down to its starting intrinsic Si substrate and characterize high-impedance ($\Zr = \SI{3}{\kilo\ohm}$) Al/AlOx/Al JJ array resonators and $\SI{50}{\ohm}$ Nb CPW resonators fabricated on the etched surface.
While the characterized JJ array resonators on the bare heterostructure exhibit $\qi \approx 3\,000-4\,000$ in the low-photon regime,  mainly due to the defects residing in various layers of the heterostructure \cite{scappucci_germanium_2021}, we find that the etching of the heterostructure down to the the Si substrate significantly enhances the internal quality factor up to $\qi \approx 10\,000 - 20\,000$. We show that both the etching process itself and the small overlapping region with the mesa do not introduce notable losses.
We further demonstrate $\qi \approx 100\,000$ for the $\SI{50}{\ohm}$ Nb resonators, confirming that the extracted $\qi$ of the JJ array resonators are limited by the JJ fabrication process.
In addition, we demonstrate the resonance frequency tunability of the JJ array resonators in a wide range of $7-\SI{4}{GHz}$ by means of an in-plane magnetic field up to $\qty{60}{\milli\tesla}$, preserving $\qi$ above $5\,000$ below $\qty{40}{\milli\tesla}$. These magnetic fields are compatible with the operation of spin qubits on planar Ge \cite{hendrickx_sweet-spot_2024, jirovec_singlet-triplet_2021, zhang_universal_2025}.
%Additionally, it is possible to increase the in-plane magnetic field resilience of JJs up to $B^\parallel_\mathrm{crit} \approx \qty{1}{T}$ by reducing the thickness of the Al leads\cite{krause_magnetic_2022}.

While we mainly focus on the resonators defined on a Ge/SiGe heterostructure, which is a promising platform for various hybrid quantum systems including hole QDs \cite{de_palma_strong_2024, janik_strong_2025}, ABSs \cite{hinderling_direct_2024, pita-vidal_direct_2023, pita-vidal_strong_2024} and minimal Kitaev chains \cite{dvir_realization_2023}, the presented tapered etching technique is a versatile approach for any kind of heterostructure comprising defects located at different layers.
Therefore, our work provides a path toward highly coherent light-matter interface in semiconductor heterostructures. 
%\cite{burkard_superconductorsemiconductor_2020}. 
%unlocking the possibilities for all types of quantum operation schemes requiring minimal loss rate of the hybrid system, namely including fast quantum state detection \cite{zheng_rapid_2019}, high-fidelity long-range interaction between remote QD states \cite{dijkema_cavity-mediated_2025, harvey-collard_coherent_2022} and microwave optics \cite{oppliger_high-efficiency_2025, liu_semiconductor_2015}.

\begin{acknowledgements}

P.S. acknowledges support from the Swiss State Secretariat for Education, Research and Innovation (SERI) under contract number MB22.00081~/~REF-1131-52105, and the NCCR SPIN, a National Centre of Competence in Research, funded by the Swiss National Science Foundation (SNSF) with grant number 225153.
P.S. also acknowledges support from the SNSF through the grants Ref. No. 200021\_200418~/~1 and Ref. No. 206021\_205335~/~1.
W.J. acknowledges support from EPFL QSE Postdoctoral Fellowship Grant.
The authors furthermore thank Camille Roy, Filippo Ferrari and Léo Peyruchat for the useful discussions. 
\section*{Authors contributions}
P.S. conceived the project. F.D.P, E.A. and R.K. developed the fabrication recipes. F.D.P. and E.A. fabricated the devices measured in this work. F.D.P., F.O. and W.J. built the experimental setup. F.D.P. and E.A. designed the devices and performed the electrical measurements. E.A. analyzed the data and prepared the data visualization. A.N and I.Z. designed and performed the growth of the SiGe heterostructure. F.D.P. performed the electromagnetic simulations. F.D.P., E.A., W.J. and P.S. wrote the manuscripts with inputs from all authors. P.S. initiated and supervised the project. F.D.P. and E.A. contributed equally to the work.
\end{acknowledgements}

\newpage

\appendix
\renewcommand{\thesubsection}{\thesection.\arabic{subsection}}
\makeatletter
\renewcommand{\p@subsection}{}

% \newcommand{\wj}[1]{\textcolor{blue}{WJ: #1}}
% \newcommand{\ea}[1]{\textcolor{magenta}{EA: #1}}
% \newcommand{\fdp}[1]{\textcolor{ForestGreen}{FDP: #1}}

% \newcommand{\qi}{Q_\mathrm{i}} % Internal quality factor inside math mode
% \newcommand{\qc}{Q_\mathrm{c}} % External quality factor inside math mode
% \newcommand{\kint}{\kappa_\mathrm{i}} % Internal loss rate
% \newcommand{\kc}{\kappa_\mathrm{c}} % Feedline-resonator coupling rate
% \newcommand{\kL}{\kappa_\mathrm{L}} % 
% \newcommand{\fd}{f_\mathrm{d}} % feedline drive frequency
% \newcommand{\omegad}{\omega_\mathrm{d}} % feedline drive frequency
% \newcommand{\Pd}{P_\mathrm{d}} % feedline drive power
% \newcommand{\fr}{f_\mathrm{r}} % Resonance frequency
% \newcommand{\tr}{S_{21}} % Transmission
% \newcommand{\Zr}{Z_\mathrm{r}} % Transmission
% \newcommand{\nph}{\langle N_\mathrm{ph} \rangle}
% \newcommand{\Bpar}{B_\parallel}

% \DeclareSIUnit{\dbm}{dBm}

\section{Fabrication methods}
\subsection{Heterostructure growth}
\label{supp:growth}
The Ge/SiGe heterostructure shown in Fig.~\ref{fig:devices}a is grown via chemical vapor deposition (CVD) by adopting a reverse-grading approach as described in ref \cite{nigro_demonstration_2024}. The full stack is $\approx \SI{1.6}{\micro\meter}$ thick, grown on a \qty{280}{\micro\meter} thick intrinsic Si wafer with a resistivity $ > \SI{10}{\kilo\ohm\cdot\cm}$. The $\SI{16}{\nano\meter}$-thick QW is buried below $\approx\SI{60}{\nano\meter}$ of SiGe spacer, terminated by $\approx \SI{2}{\nano\meter}$ of Si cap, fully oxidized by an additional oxygen plasma treatment \emph{ex-situ} just before starting the fabrication process.

\subsection{Tapered heterostructure etching}
\label{supp:etching}
To perform the tapered Ge/SiGe heterostructure etching, the sample is spin-coated with \qty{2}{\micro\meter} of positive-tone photoresist AZ10XT-20. The etching mask is patterned with photolithography using a \qty{405}{\nano\meter} direct laser writer. In order to provide the mesa step with a tapered profile, the sample is baked at \qty{135}{\celsius} for 2 minutes: in this way, the resist reflows and its sidewalls acquire a rounded profile, which imprints a tapered edge to the heterostructure during the etching. The heterostructure is etched using an inductively coupled plasma (ICP) dry etching process using Cl$_2$ chemistry. During the etching process, the chamber is at a pressure of \qty{5}{mTorr}, the Cl$_2$ gas flow is kept at \qty{50}{sccm}, the RF power to generate the plasma is \qty{800}{W} and the RF power to accelerate the ions and control their mean energy is \qty{100}{W}. A laser-based end point detection system reveals when the Si substrate is reached and is used to stop the etching process without excessively over-etching the Si substrate. Immediately after the etching process, the sample is rapidly neutralized from Cl$_2$ in H$_2$O, and the etching mask stripped in a solution of 1-Methyl-2-pyrrolidinone (NMP) at \qty{70}{\celsius} for a couple of hours. 

\subsection{Josephson junction array resonators}
\label{supp:fab_JJ_arrays}
The fabrication of the Josephson junction array resonators consists of two steps: (a) Nb ground plane deposition and (b) Al-AlOx-Al Josephson junction deposition. 

The ground plane is patterned using photolithography and lift-off; the sample is spin-coated with \qty{400}{nm} of negative-tone resist LOR 5A followed by \qty{1.1}{\micro\meter} of positive-tone resist AZ 1512 HS and patterned using a \qty{405}{\nano\meter} direct laser writer. \qty{30}{nm} of Nb are deposited in a Plassys UHV e-beam evaporator at a rate of \qty{0.5}{nm \per\second}. Without breaking the vacuum, Nb is oxidized statically at a pressure of \qty{10}{torr} for 10 minutes to provide a first clean oxidized interface. Nb is lifted off in an NMP solution at \qty{70}{\celsius} for a couple of hours. 

To pattern the Dolan-style JJ array resonators the sample is spin-coated with \qty{400}{nm} of MMA followed by \qty{220}{nm} of PMMA. The resonator body and the junctions' suspended bridges are patterned with 100 keV electron-beam lithography. The junctions are deposited in a Plassys UHV e-beam evaporator; a first layer of \qty{35}{nm} of Al is deposited with an angle of +\ang{45} at a rate of \qty{0.5}{nm \per\second}. Without breaking the vacuum, the sample is oxidized statically at a pressure of \qty{2}{torr} for 20 minutes. A second layer of \qty{130}{nm} of Al is deposited with an angle of $-\ang{45}$ at a rate of \qty{0.5}{nm \per\second}. Finally, the sample is oxidized again at a pressure of \qty{10}{torr} for 10 minutes to provide a first clean oxidized interface. Al is lifted off in an NMP solution at \qty{70}{\celsius} for a couple of hours. 

\subsection{CPW Nb resonators}
\label{supp:fab_CPW}
The CPW Nb resonators are patterned using photolithography and lift-off. The sample is spin-coated with \qty{400}{nm} of LOR 5A resist followed by \qty{1.1}{\micro\meter} of AZ 1512 HS resist and patterned using a \qty{405}{\nano\meter} direct laser writer. \qty{50}{nm} of Nb are deposited in a Plassys UHV e-beam evaporator at a rate of \qty{1}{nm \per\second}. The sample is taken out of the evaporator, oxidized in air and lifted off in an NMP solution at \qty{70}{\celsius} for a couple of hours.

%%%%%%%%%%%%%%%%%%%%%%%%%%%%%%%%%%%%%%%%%%%%%%%%%%%%%%%%%%%%%%%%%%%%%%%%%%%%%%%%%%%%%%%%%%%%%%
\section{Measurement setup}
\label{supp:measurement_setup}
The measurements reported in this work are performed in a dilution refrigerator (Bluefors LD400) equipped with a 6-1-1 Tesla vector magnet (American Magnetics), at base temperatures around 10 mK.
The devices are mounted on a printed circuit board (PCB) hosting RF transmission lines with a characteristic impedance of $Z_0 = \SI{50}{\ohm}$. The transmission lines are wire-bonded to the on-chip photon feedlines on one end, and directly connected to coaxial cables for RF signals on the other end.   
Resonator spectroscopy is performed with a vector network analyzer (VNA, Rohde \& Schwarz ZNB20). The VNA output is attenuated by \qty{20}{dB} at room temperature, followed by a DC block (Inmet 8039 inner-outer). The signal is then attenuated at various stages in the cryostat with a total attenuation of \qty{60}{dB}. In addition, from previous line calibration measurements we estimated an additional attenuation of the signal by $\approx \qty{10}{dB}$ before reaching the device, due to cable losses. The signal transmitted through the device then passes through two isolators (Low Noise Factory ISISC4\_8A) and two circulators (Low Noise Factory CICIC4\_8A). The signal is amplified by a HEMT at the \qty{4}{\kelvin} stage (Low Noise Factory LNC4\_8C) and by a low noise amplifier at room temperature (Agile AMT A0284) after passing through another dc block (Inmet 8039).

%%%%%%%%%%%%%%%%%%%%%%%%%%%%%%%%%%%%%%%%%%%%%%%%%%%%%%%%%%%%%%%%%%%%%%%%%%%%%%%%%%%%%%%%%%%%%%
\section{Resonator design and parameters}
\subsection{Josephson junction array resonators}
\label{supp:design_JJ_arrays}
From room temperature resistance measurements of junction test structures, we extract a resistance per JJ of $R_\mathrm{JJ} \approx \SI{1.25}{\kilo\ohm}$, corresponding to an inductance per junction of $L_\mathrm{JJ} \approx \SI{1.48}{\nano\henry}$, a critical current of $I_\mathrm{c,JJ} \approx \SI{220}{\nano\ampere}$ and a Josephson energy $E_\mathrm{J} \approx \qty{111}{\giga\hertz}$. From SEM measurements, we get a JJ area of $w \times l = \qtyproduct{520 x 760}{\nano\meter}$; assuming a thickness $t_\mathrm{ox} \approx \qty{1}{nm}$ for the oxide tunneling barrier, we compute the junction's self-capacitance $C_\mathrm{J} \approx \qty{31.5}{\femto\farad}$, corresponding to a charging energy $E_\mathrm{C} \approx \qty{0.6}{\giga\hertz}$. We obtain a ratio $E_\mathrm{J}/E_\mathrm{C} \approx \qty{180}{}$ and a plasma frequency $\omega_\mathrm{p} \approx \qty{23}{\giga\hertz}$. 

With a $\qty{40}{\micro\meter}$ spacing of the JJ array resonator to the ground plane, we estimate a capacitance per unit length to ground of $C_\mathrm{0,Si} \approx \qty{0.057}{\femto\farad/\micro\meter}$ ($C_\mathrm{0,SiGe} \approx \qty{0.061}{\femto\farad/\micro\meter}$) on the Si (SiGe heterostructure) substrate \cite{goppl_coplanar_2008}. The difference in capacitance originates from the difference in dielectric constant of the two substrates, respectively assumed to be 11.9 and 13.0. Concerning the central JJ array resonator of the five, we concatenate $N=46$ junctions. We take into account the aforementioned junctions, as well as the extra spurious junctions originated by the Dolan-bridge technique ($\approx \qty{3.5}{\micro\meter}$ long, hence contributing much less to the total inductance), and we obtain a total inductance of the array $L_\mathrm{tot} \approx \qty{78.9}{\nano\henry}$ and a total length of $\qty{207}{\micro\meter}$. We treat the resonator as a distributed element \cite{castellanos-beltran_widely_2007}. By grounding one resonator's side while keeping the other end open (see Fig.\ref{fig:devices}e), we define a quarter-wave fundamental mode, with a lumped equivalent inductance of $L_\mathrm{eq} \approx \qty{67}{\nano\henry}$ and capacitance $C_\mathrm{eq,Si} \approx \qty{5.78}{\femto\farad}$ ($C_\mathrm{eq,SiGe} \approx \qty{6.27}{\femto\farad}$) \cite{goppl_coplanar_2008}. These values result in a bare resonance frequency of $f_\mathrm{Si} \approx \qty{8.09}{\giga\hertz}$ ($f_\mathrm{SiGe} \approx \qty{7.77}{\giga\hertz}$). 

We add the coupling hook shown in Fig.\ref{fig:devices}e (green color) to determine the external coupling rate $\kc$ to the photon feedline. For the resonators on the bare SiGe heterostructure we design a coupling hook that is longer than on Si ($\qty{100}{\micro\meter}$-long vs $\qty{40}{\micro\meter}$-long), in order to provide a higher coupling rate. At the voltage anti-node of the resonator we add the $\qtyproduct{150 x 150}{\micro\meter}$ mesa island, together with a $\qtyproduct{2 x 10}{\micro\meter}$ patch of the resonator end that climbs over the mesa step (see Fig.\ref{fig:devices}f), overlapping with the mesa with an area of $A_\mathrm{overlap} = \qtyproduct{2 x 5}{\micro\meter}$. As explained in the main text, this patch is realized in the same evaporation step of the JJs, therefore its inductance can be neglected. Although the resonators on intrinsic Si and on SiGe do not need to climb any mesa, we also design them with a patch a the voltage antinode with the same dimensions, in order to compare resonators with the same design.

With this JJ array resonator design, we perform electromagnetic (EM) simulations (Sonnet software) to properly estimate both the coupling quality factor and the loaded resonance frequency. The simulation setup is shown in Fig.\ref{fig:supp_simulation_schematic}. To do so, we replace the JJ array with a strip of the same length and width, and with a sheet inductance of $L_\mathrm{\square} \approx \qty{208}{\pico\henry}$/$\square$. At this simulation stage, the substrate losses that we provide in the simulation software can be arbitrarily chosen.
%We remind that the resonators fabricated and characterized in this work are designed with a patch of $\qtyproduct{2 x 10}{\micro\meter}$, . For the simulation results of the $A_\mathrm{overlap}$ sweep, please refer to Appendix~\ref{}. 
Finally, we fit the simulation results to Eq.~\ref{eq:S21_linear} and we extract a resonance frequency of $f_\mathrm{Si} \approx \qty{7.02}{\giga\hertz}$ ($f_\mathrm{SiGe} \approx \qty{6.61}{\giga\hertz}$) and a coupling quality factor of $Q_\mathrm{c,Si} \approx 1\,000$ ($Q_\mathrm{c,SiGe} \approx 600$). From the simulated resonance frequency we extract the total $C_\mathrm{eq}$, keeping the same $L_\mathrm{eq}$ previously computed. This allows us to estimate an equivalent impedance of the resonator $Z_\mathrm{eq} = \qty{3}{\kilo\ohm}$ for both substrates. In conclusion, to design the remaining four resonators hanged to the same transmission line, we just sweep the total number of junctions from $N=42$ to $N=50$ to effectively get a frequency spacing of $\approx \qty{200}{\mega\hertz}$ between each resonator.
For EM simulations that take into account the losses originating from the mesa climb on the etched heterostructure, see Appendix~\ref{supp:simulations_losses}.

%%%%%%%%%%%%%%%%%%%%%%%%%%%%%%%%%%%%%%%%%%%%%%%%%%%%%%%%%%%%%%%%%%%%%%%%%%%%%%%%%%%%%%%%%%%%%%
\subsection{\qty{50}{\ohm} CPW resonators}
\label{supp:design_CPW}
For the Nb CPW resonators we follow a standard resonator design, where the resonance frequency and impedance depend both on geometric parameters only. The inner conductor is \qty{16}{\micro\meter} wide and the gap between the inner conductor and the ground plane is \qty{10}{\micro\meter} wide, resulting in a characteristic impedance of \qty{50}{\ohm} on a Si substrate. The length of the resonators is adjusted to target resonance frequencies of $\qtyrange[range-units = single, range-phrase = -]{5}{6}{\GHz}$. The resonators are open on one side and shunted to the ground plane on the other side, in a quarter-wave geometry. The very high coupling quality factors $Q_\mathrm{c} \approx 100\,000$ are achieved just by placing the current anti-node of the resonator few $\qty{}{\micro\meter}$ far away from the feedline, as shown in Fig.\ref{fig:cpw}b.

\section{Electromagnetic simulations for losses estimations}
\label{supp:simulations_losses}
\begin{figure*}[!ht]
    \centering
    \includegraphics{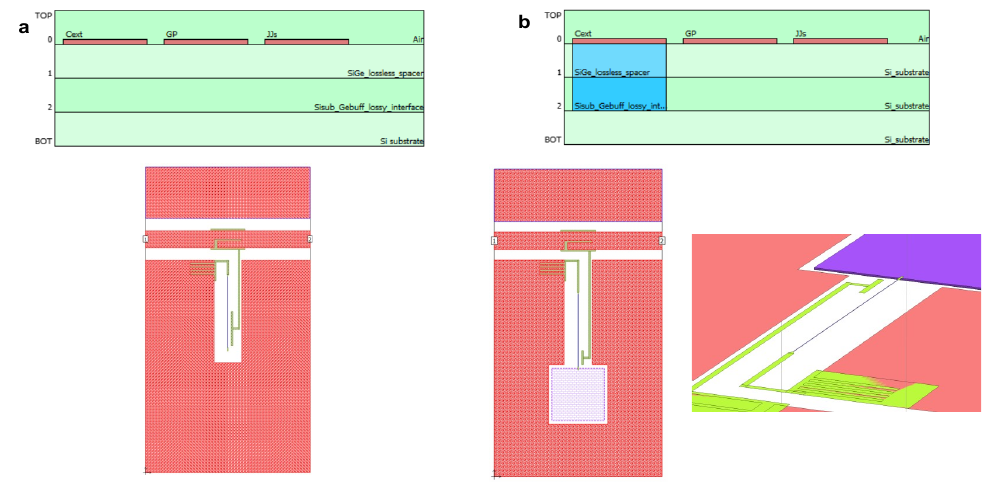}
    \caption{Sonnet simulation setup. \textbf{\textsf{(a)}} Cross section and top view of a resonator on top of the bare SiGe heterostructure. \textbf{\textsf{(b)}} Cross section, top view and tilted view of a resonator on top of the Si substrate and climbing the SiGe mesa.}
    \label{fig:supp_simulation_schematic}
\end{figure*}
To further validate the efficacy of our tapered etching approach, we perform electromagnetic (EM) simulations using Sonnet\textregistered. As outlined in the main text, the primary source of loss in reverse-graded Ge/SiGe heterostructures originates at the lattice-mismatched interface between the initial Si wafer and the Ge virtual substrate. Moreover, as discussed in ref \cite{nigro_demonstration_2024}, the independence of the $\qi$ on $\nph$ suggests that the loss mechanism cannot be saturated with the power. For this reason, we treat microwave losses as resistive losses as a result of the coupling of the resonator to a conductive layer buried at the interface between the Si substrate and the Ge virtual substrate. We define this layer $\SI{100}{\nano\meter}$ thick and encapsulate it between the intrinsic (conductivity at $\SI{10}{\milli\kelvin}$: $\sigma=\SI{4.4e-7}{\siemens/\meter}$) Si substrate (dielectric losses: loss tangent tan$\delta=\SI{1.0e-5}{}$) and a \qty{1.5}{\micro\meter} thick intrinsic (conductivity at $\SI{10}{\milli\kelvin}$: $\sigma=\SI{4.4e-7}{\siemens/\meter}$) SiGe spacer (dielectric losses: loss tangent tan$\delta=\SI{1.0e-5}{}$). The total thickness of the full stack is $\SI{1.6}{\micro\meter}$, as shown in Fig.\ref{fig:supp_simulation_schematic}. The assumption of identical losses for both the Si substrate and the SiGe spacer is unrealistic, given the presence of TLFs in the SiGe heterostructure. However, as the predominant losses arise from the 100 nm lattice-mismatched layer, a precise quantification of the TLF losses in the SiGe layer falls beyond the scope of this work. Please note that the thickness of $\qty{100}{\nano\meter}$ is completely arbitrary: it represents just a way to confine the losses to a buried layer not in direct galvanic contact with the resonator. On the other hand, a loss tangent tan$\delta=\SI{1.0e-5}{}$ establishes a more realistic upper bound for $\qi \approx 100\,000$. Without appropriate surface treatment to remove surface contaminants and native SiO$_\mathrm{2}$, achieving larger $\qi$ would not be feasible in the low photon regime, even for simple $\qty{50}{\Omega}$ Nb resonators. In fact, such aggressive surface treatments are typically avoided in QW-based semiconductor devices.

\begin{figure}
    \centering
    \includegraphics{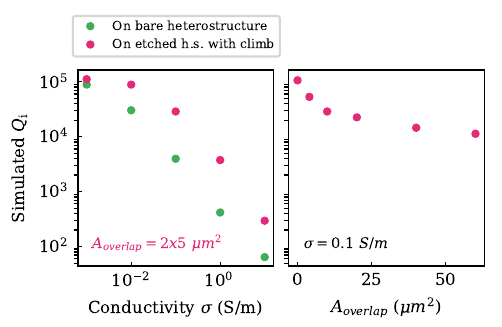}
    \caption{Results of the simulated $\qi$. \textbf{\textsf{(a)}} $\qi$ as a function of conductivity $\sigma$ for both the bare heterostructure and the etched down to the Si substrate cases. Here, the overlap area with the mesa is fixed to $A_\mathrm{overlap} = \qtyproduct{2 x 5}{\micro\meter}$, as in the experiments. \textbf{\textsf{(b)}} $\qi$ as a function of $A_\mathrm{overlap}$ for a fixed $\sigma = \qty{0.1}{\siemens/\meter}$.}
    \label{fig:supp_Qi_vs_cond_vs_area}
\end{figure}

In all the simulation results shown in this section, the object of the study is the same resonator described in Appendix~\ref{supp:design_JJ_arrays}. In the first set of simulations, shown in Fig.~\ref{fig:supp_Qi_vs_cond_vs_area}a, we study $\qi$ as a function of the conductivity $\sigma$ of the $\qty{100}{\nano\meter}$ lattice-mismatched layer. We sweep $\sigma$ by five orders of magnitude, from $\qty{0.001}{\siemens/\meter}$ to $\qty{10}{\siemens/\meter}$, corresponding to a resistivity $\rho$ varying from $\qty{100}{\kilo\Omega\cdot\centi\meter}$ to $\qty{10}{\Omega\cdot\centi\meter}$. Fig.~\ref{fig:supp_Qi_vs_cond_vs_area}a reports the resulting $\qi$ as a function of $\sigma$ in two different cases. In the first case (green dots), the resonator lies on top of the full heterostructure, while in the second case (pink dots) the body of the resonator lies on top of the Si substrate, and climbs a $\qtyproduct{150 x 150}{\micro\meter}$ mesa only with an overlap area of $A_\mathrm{overlap} = \qtyproduct{2 x 5}{\micro\meter}$, exactly as in the case of the fabricated and characterized devices. With respect to the bare heterostructure case, the simulation predicts an increase of about one order of magnitude in $\qi$ for $\sigma \geq \qty{0.1}{\siemens/\meter}$ with our etching approach. This suggests that the proposed tapered etching approach can also be efficient for resonators directly fabricated on Ge/SiGe heterostructures which present $\qi < 1\,000$. We note that the case $\sigma = \qty{0.1}{\siemens/\meter}$ predicts $\qi \approx 4\,000$ on the bare heterostructure, and $\qi \approx 30\,000$ on the etched one. This scenario may reproduce our experimental data for the JJ array resonator case. Unfortunately, as discussed in the main text, the quality of the JJ fabrication does not allow us to observe $\qi$ beyond $10\,000-20\,000$. 

In the second simulation, whose results are reported in Fig. \ref{fig:supp_Qi_vs_cond_vs_area}b, we fix $\sigma = \qty{0.1}{\siemens/\meter}$ and sweep the area of the overlap region from $\qty{0}{}$ to $\qtyproduct{2 x 30}{\micro\meter}$. Within this interval, we notice a drop in $\qi$ by one order of magnitude (from $ \approx 100\,000$ to $ \approx 10\,000$), which underlines the importance of keeping the overlap area as small as possible in order to minimize the losses.

%%%%%%%%%%%%%%%%%%%%%%%%%%%%%%%%%%%%%%%%%%%%%%%%%%%%%%%%%%%%%%%%%%%%%%%%%%%%%%%%%%%%%%%%%%%%%%
\section{Resonator fitting models}

\subsection{Linear model}
\label{supp:fitting_linear}
The experimental spectroscopy data $\tr (\fd)$ is fitted to a master equation model. For the JJ array resonators in the low-photon regime and for the CPW resonators at any photon number, we describe the resonators with a linear Hamiltonian:
\begin{equation}
    \mathcal{H} = \hbar \omega_0 a^\dagger a.
\end{equation}
We follow the standard input-output theory\cite{gardiner_input_1985,chen_scattering_2022} to write the Heisenberg-Langevin equation of motion for the $a$ field operator for a notch-type resonator in the rotating frame at $\omegad = 2\pi \fd$:
\begin{equation}
    \dot{a} = -i (\omega_0 - \omega_d) a - \frac{\kc + \kint}{2} a - \sqrt{\frac{\kc}{2}} b_\mathrm{in},
    \label{eq:Langevin}
\end{equation}
where $b_\mathrm{in}$ represents the input photon field. In the following, we use $\Delta_\mathrm{r} = \omega_0 - \omega_d$. Stationary solutions for Eq.~\ref{eq:Langevin} are found by imposing $\dot{a} = 0$, leading to:
\begin{equation}
    a = - \frac{\sqrt{\kc/2}}{\frac{\kc + \kint}{2} + i \Delta_\mathrm{r}} b_\mathrm{in}.
    \label{eq:stationary_Langevin}
\end{equation}
The relation between the input photon field $b_\mathrm{in}$ and the output photon field $b_\mathrm{out}$ reads:
\begin{equation}
    b_\mathrm{out} = b_\mathrm{in} + \sqrt{\frac{\kc}{2}}a.
    \label{eq:IO}
\end{equation}
Combining Eq.~\ref{eq:stationary_Langevin} and Eq.~\ref{eq:IO} we obtain:
\begin{equation}
    \tr = \frac{\langle b_\mathrm{in}\rangle}{\langle b_\mathrm{out}\rangle} = 1 - \frac{\kc /2}{\frac{\kc + \kint}{2} + i \Delta_\mathrm{r}}.
\end{equation}
If we add to this simple model the presence of a non-ideal response of the cavity due to environmental factors, we obtain the following model\cite{probst_efficient_2015} which we use for fitting the resonators in the linear regime:
\begin{equation}
    \tr = a e^{i\alpha} e^{- i \omegad \tau} \frac{2 i\Delta_\mathrm{r} + \kL - \kc e^{i\phi_0}/\cos(\phi_0)}{2i \Delta_\mathrm{r} + \kL},
    \label{eq:S21_linear}
\end{equation}
where we have defined the loaded quality factor $\kL = \kint + \kc$ and $a$, $\alpha$, $\tau$ and $\phi_0$ are correction factors to take into account the non-ideal response of the cavity due to the environment. 
From Eq.~\ref{eq:stationary_Langevin} we calculate the number of photons\cite{eichler_controlling_2014} at $\omegad = \omega_0$:
\begin{equation}
    \nph = \langle a^\dagger a\rangle = \frac{2 \kc}{\kappa_\mathrm{L}^2} \langle b_\mathrm{in}^\dagger b_\mathrm{in} \rangle = \frac{2 \kc}{\kappa_\mathrm{L}^2} \frac{P[\mathrm{W}]}{\hbar \omega_0}.
\end{equation}
\subsection{Nonlinear model}
\label{supp:fitting_nonlinear}
For a nonlinear model that takes into account the Kerr nonlinearity of the JJ array, we describe the resonator with the Hamiltonian\cite{eichler_controlling_2014}:
\begin{equation}
    \mathcal{H} = \hbar \omega_0 a^\dagger a - \hbar \frac{K}{2} a^\dagger a a^\dagger a.
\end{equation}
Following the same input-output theory derivation described above, we obtain the following scattering parameter\cite{eichler_controlling_2014, roy_study_2025}:
\begin{equation}
    \tr = a e^{i\alpha} e^{- i \omegad \tau} \left( 1 - \frac{\kc}{\kc + \kint} \frac{e^{i\phi}}{\cos(\phi)}\frac{1}{1 + 2i (\delta - \xi n)} \right),
    \label{eq:S21_nonlinear}
\end{equation}
where:
\begin{equation}
    \delta = \frac{\omegad - \omega_0}{\kc + \kint}, \quad \xi = \frac{|\alpha_\mathrm{in}|^2 \kc K}{(\kc + \kint)^3}, \quad n = \frac{\nph}{|\alpha_\mathrm{in}|^2}\frac{(\kc + \kint)^2}{\kc}.
\end{equation}
$|\alpha_\mathrm{in}|^2 = P[\mathrm{W}]/\hbar \omegad$ is the input photon flux and $K$ is the self-Kerr parameter. The renormalized number of photons $n$ is calculated by solving the equation:
\begin{equation}
    \frac{1}{2} = \left( \delta^2 + \frac{1}{4} \right) n - 2\delta \xi n^2 + \xi^2 n^3.
\end{equation}
To fit the full power sweep spectroscopy data of JJ array resonators like the one represented in Fig.~ \ref{fig:resonances}d, we first perform a linear fit of single power slices of the data to Eq.~\ref{eq:S21_linear} below the single photon power, in order to extract the parameters $\omega_0$, $\kc$, $\kint$ and the environmental parameters. We then fit the whole 2D data of the power sweep spectroscopy to Eq.~\ref{eq:S21_nonlinear}, optimizing only the values of $K$ and $\phi$ while keeping the values of $\omegad$, $\kc$ and $\kint$ fixed to the values obtained from the linear fit. We operate in this way because at high photon number the fitting procedure typically struggles to properly extract all the fitting parameters independently, and our goal for the nonlinear fit is to estimate the self-Kerr nonlinearity $K$.

\section{Power spectroscopies for all resonators}
\label{supp:power_sweeps_all}
\begin{figure*}[!ht]
    \centering
    \includegraphics{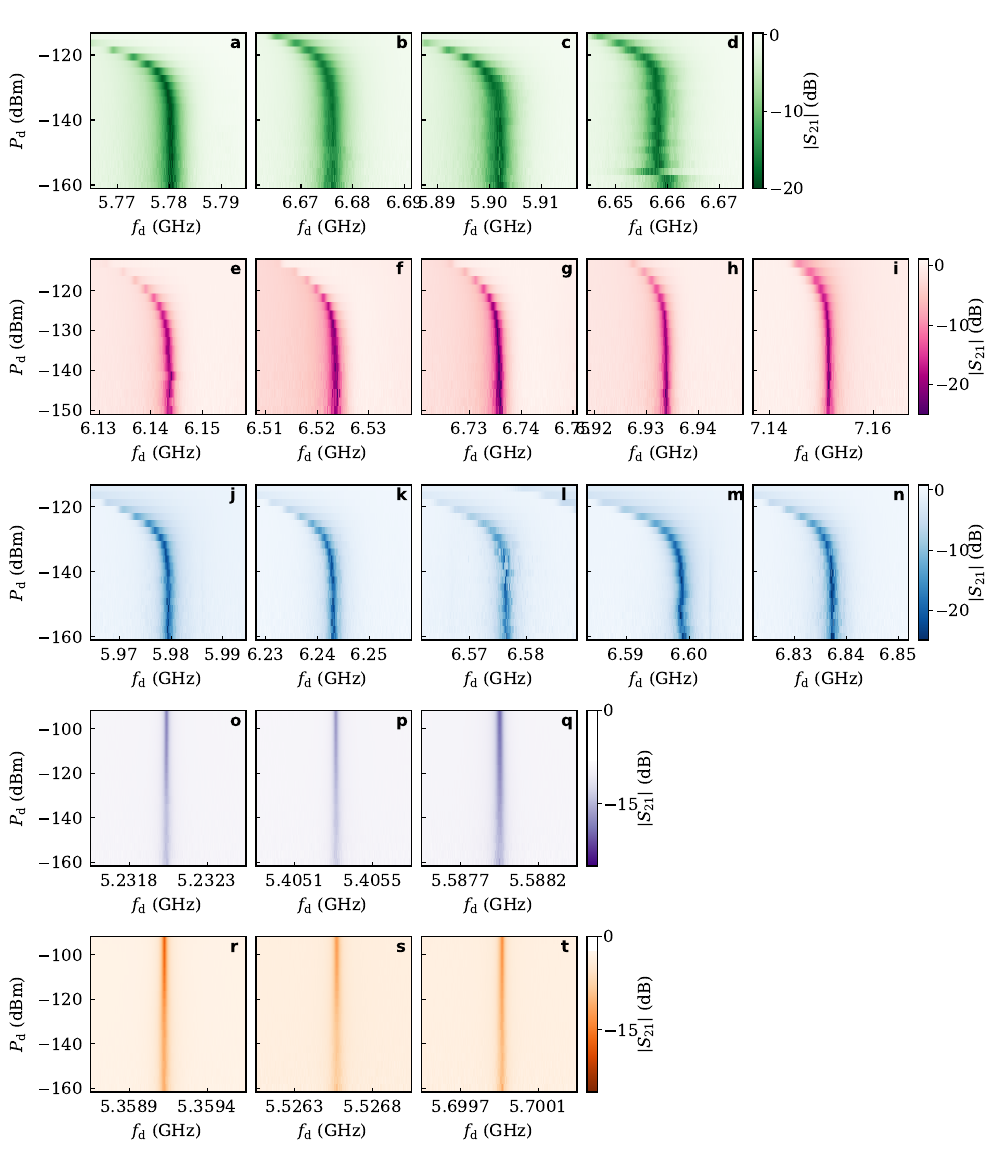}
    \caption{Power spectroscopies for all the measured resonators. Consistently with the main text, green color (\textbf{\textsf{(a)-(d)}}) is used for JJ arrays on the bare heterostructure; pink color (\textbf{\textsf{(e)-(i)}}) is used for JJ arrays on the etched heterostructure with climb; blue color (\textbf{\textsf{(j)-(n)}}) is used for JJ arrays on the Si reference substrate; purple color (\textbf{\textsf{(o)-(q)}}) is used for CPW resonators on the etched heterostructure; orange color (\textbf{\textsf{(r)-(t)}}) is used for CPW resonators on the Si reference substrate.}
    \label{fig:power_sweep_all}
\end{figure*}

    Figure~\ref{fig:power_sweep_all} shows power spectroscopy measurements for all resonators. By fitting these power scans to Eq.~\ref{eq:S21_linear} in the low-photon-number regime, we obtain the internal quality factors shown in Figs.~\ref{fig:resonances}l, ~\ref{fig:cpw}c. We note that for some resonators it is possible to observe the strong hybridization, and the corresponding resonance splitting, of the resonators with TLFs that reside in the junctions' oxide tunneling barriers (see for instance Figs.~\ref{fig:power_sweep_all}d,~l,~m). By fitting the full power sweeps of the JJ arrays to the nonlinear model in Eq.~\ref{eq:S21_nonlinear}, we consistently find Kerr nonlinearities in the range $K \approx \qtyrange[range-phrase=~-~]{100}{300}{kHz}$. These values are consistent with an approximate theoretical estimate of the JJ array's nonlinearity: for a single JJ, indeed, $K_1 = E_C \approx \qty{600}{MHz}$ \cite{blais_circuit_2021} (see Appendix~\ref{supp:design_JJ_arrays}); given that the Kerr nonlinearity scales\cite{eichler_controlling_2014} with the number of junctions $N$ approximately as $K_N \propto 1/N^2$, we obtain $K_N \approx E_C/N^2 \approx \qty{280}{kHz}$ for $N = 46$.

\section{Extracted coupling quality factors}
\label{supp:Qc_vs_Nph}
\begin{figure}[h!]
    \centering
    \includegraphics{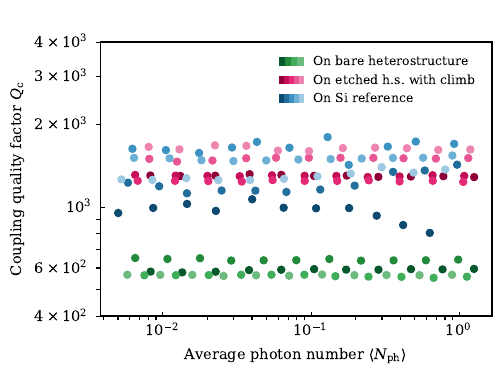}
    \caption{Extracted coupling quality factors for the different JJ array resonators on the different substrates. The extracted $\qi$ are reported in Fig.~\ref{fig:resonances}l.}
    \label{fig:supp_Qc_vs_Nph}
\end{figure}

\begin{figure}[h!]
    \centering
    \includegraphics{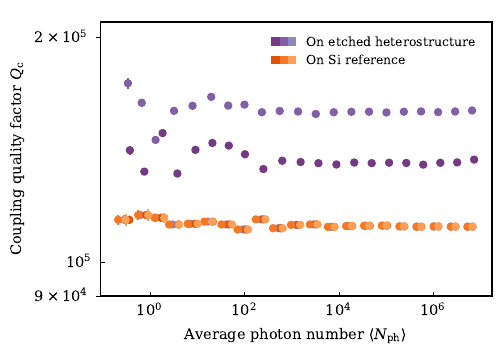}
    \caption{Extracted coupling quality factors for the different CPW Nb array resonators on the different substrates. The extracted $\qi$ are reported in Fig.~\ref{fig:cpw}c.}
    \label{fig:supp_Qc_vs_Nph_CPW}
\end{figure}

In Fig.\ref{fig:supp_Qc_vs_Nph} (Fig.\ref{fig:supp_Qc_vs_Nph_CPW}) we report the extracted coupling quality factors $Q_\mathrm{c}$ of the JJ array (the Nb $\qty{50}{\ohm}$) resonators. $Q_\mathrm{c}$ is $\approx 1\,000-2\,000$ for the JJ array resonators on the Si reference and on the etched heterostructure, and $\approx 600$ for the ones on top of the bare heterostructure, matching the simulated ones (see Appendix~\ref{supp:design_JJ_arrays}). For the $\qty{50}{\ohm}$ resonators, we extract  $Q_\mathrm{c}$ of $100\,000-200\,000$, also in this case perfectly matching the simulated ones (see Appendix~\ref{supp:design_CPW}).

\section{In-plane magnetic field dependence of JJ arrays' resonance frequencies}
\label{supp:B_field_behaviour}
The resonance frequency of Josephson junction array resonators depends on the presence of an external magnetic field. In this work, we investigated this behavior in the presence of an in-plane magnetic field $B_\parallel$ applied parallel to the JJ array's axis. The dependence of the resonance frequency $\fr$ on $B_\parallel$ is due to the interplay of two effects.
Firstly, an external magnetic field reduces the superconducting gap of the Al leads of the JJs according to\cite{tinkham_introduction_1996, krause_magnetic_2022}:
\begin{equation}
    \Delta(B_\parallel) = \Delta_0 \sqrt{1 - \left( \frac{B_\parallel}{B_\mathrm{crit}^\parallel} \right)^2},
    \label{eq:Delta_Bin}
\end{equation}
where $B_\mathrm{crit}^\parallel$ is the in-plane critical field of the Al leads and $\Delta_0$ is the zero-field superconducting gap of the Al leads. Notably, $B_\mathrm{crit}^\parallel$ depends on the thickness of the Al leads. For a thin superconducting film of thickness $d$, $B_\mathrm{crit}^\parallel$ is \cite{tinkham_introduction_1996,krause_magnetic_2022}:
\begin{equation}
    B_\mathrm{crit}^\parallel = B_\mathrm{crit}^\mathrm{bulk} \frac{\lambda_\mathrm{eff}}{d} \sqrt{24},
    \label{eq:B_crit}
\end{equation}
where $B_\mathrm{crit}^\mathrm{bulk}$ is the bulk critical field of the superconductor and $\lambda_\mathrm{eff}$ is the effective magnetic penetration depth of the superconductor. The latter also depends on the film thickness as\cite{krause_magnetic_2022}:
\begin{equation}
    \lambda_\mathrm{eff} = \lambda_\mathrm{L} \sqrt{1 + \frac{\xi_0}{\ell}} \approx \lambda_\mathrm{L} \sqrt{\frac{\xi_0}{d}},
    \label{eq:lambda_eff}
\end{equation}
where $\lambda_\mathrm{L}$ is the material's London penetration depth, $\xi_0$ is the material's Pippard coherence length and $\ell$ is the Cooper pairs' mean free path. The last approximation in Eq.~\ref{eq:lambda_eff} is valid if $d \ll \ell$ (and hence $\ell$ can be replaced with $d$) and $d\ll \xi_0$. For the Josephson junctions of this work, we use standard Al parameters\cite{krause_magnetic_2022} $\lambda_\mathrm{L} \approx \qty{16}{nm}$, $\xi_0 \approx \qty{1600}{nm}$ and  $B_\mathrm{crit}^\mathrm{bulk} \approx \qty{10}{mT}$ and predict $\lambda_\mathrm{eff,1} \approx \qty{129}{nm}$ and $B_\mathrm{crit,1}^\parallel \approx \qty{254}{mT}$ for the bottom Al layer ($d_1 = \qty{35}{nm}/\sqrt{2} \approx \qty{25}{nm}$), and $\lambda_\mathrm{eff,2} \approx \qty{67}{nm}$ and $B_\mathrm{crit,2}^\parallel \approx \qty{36}{mT}$ for the top Al layer ($d_2 = \qty{130}{nm}/\sqrt{2} \approx \qty{92}{nm}$). We expect the actual critical field to be dominated by the minimum between $B_\mathrm{crit,1}^\parallel$ and $B_\mathrm{crit,2}^\parallel$, hence $B_\mathrm{crit}^\parallel \approx \qty{36}{mT}$.

Eq.~\ref{eq:Delta_Bin} translates into a magnetic-field dependent critical current $I_c = \pi \Delta(B_\parallel)/(2R_N)$, where $R_N$ is the normal-state resistance of the junction, which in turn translates into a magnetic-field dependent junction inductance $L_J = \Phi_0/(2\pi I_c)$. Finally, the resonance frequency of the JJ array resonators depends on the junction's inductance as $f_r \propto 1/\sqrt{L_J}$, hence:
\begin{equation}
    f_r \propto \sqrt[4]{1 - \left( \frac{B_\parallel}{B_\mathrm{crit}^\parallel} \right)^2}
    \label{eq:fr_Delta_suppression}
\end{equation}

\begin{figure}[h!]
    \centering
    \includegraphics[width=6cm]{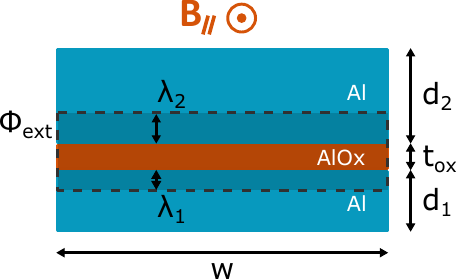}
    \caption{Schematic cross-sectional view of a JJ, with bottom and top Al leads (blue color) with thicknesses $d_1$  and $d_2$ respectively, and an oxide tunneling barrier (orange color) with thickness $t_\mathrm{ox}$. The external in-plane magnetic field (denoted by the orange circle above the junction) penetrates in the full junction width and, in the vertical direction, in the oxide tunneling barrier and in the Al leads within their effective penetration depths (area delimited by the dashed line and shaded in darker color). By penetrating in such region, it creates a magnetic flux that modulates the critical current of the JJ.}
    \label{fig:field_penetration_in_junction}
\end{figure}

In addition to that, an in-plane magnetic field can thread the junction in the region corresponding to the oxide tunneling barrier and into the Al leads within their penetration depth. In such a way, the magnetic field can effectively generate a magnetic flux $\Phi_\mathrm{ext}$, which further modulates the critical current of the junctions according to\cite{kuzmin_tuning_2023, winkel_superconducting_nodate, krause_magnetic_2022, tinkham_introduction_1996}:
\begin{equation}
    I_c \rightarrow I_c~\mathrm{sinc} \left(\pi \frac{\Phi_\mathrm{ext}}{\Phi_0}\right) = I_c~\mathrm{sinc}  \left(\pi \frac{B_\parallel}{B_{\Phi_0}}\right),
    \label{eq:I_c_fraunhofer}
\end{equation}
where $B_{\Phi_0}$ is the value of $B_\parallel$ for which $\Phi_\mathrm{ext}$ equals one superconducting flux quantum $\Phi_0 = h/(2e)$. 
This can be calculated by knowing the size of the area $A$ where $B_\parallel$ penetrates, which is approximately equal to\cite{winkel_superconducting_nodate}:
\begin{equation}
    A = w \left( \lambda_1 + t_\mathrm{ox} + \lambda_2 \right),
\end{equation}
where $\lambda_{1,2} = \mathrm{min}(\lambda_\mathrm{eff,1,2},~d_{1,2})$ is the actual portion of the Al leads where $\Bpar$ penetrates, and $t_\mathrm{ox}$ is the thickness of the oxide tunneling barrier, as schematically illustrated in Fig.~\ref{fig:field_penetration_in_junction}. $B_{\Phi_0}$ is then given by:
\begin{equation}
    B_{\Phi_0} = \frac{\Phi_0}{A}.
\end{equation}
With the values calculated above and assuming $t_\mathrm{ox} \approx \qty{1}{nm}$, we predict $B_{\Phi_0} \approx \qty{42}{mT}$.

By combining the effects described in Eqs.~\ref{eq:Delta_Bin} and \ref{eq:I_c_fraunhofer}, one obtains the following prediction for the dependence of $\fr$ on $\Bpar$:
\begin{equation}
    \fr(\Bpar) = f_0~\sqrt[4]{1 - \left( \frac{B_\parallel}{B_\mathrm{crit}^\parallel} \right)^2} \sqrt{\mathrm{sinc} \left(\pi \frac{B_\parallel}{B_{\Phi_0}}\right)},
    \label{eq:fr_of_Bin}
\end{equation}
where $f_0$ is the zero-field resonance frequency of the JJ array resonators.

\begin{figure}[h!]
    \centering
    \includegraphics{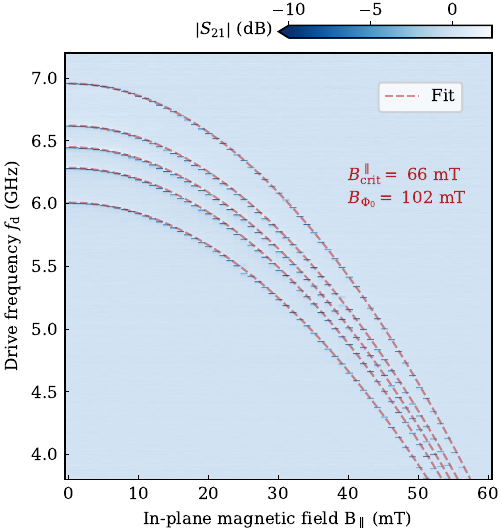}
    \caption{Fit of the magnetic field spectroscopy reported in Fig.~\ref{fig:magnetic_field}a,  according to Eq.~\ref{eq:fr_of_Bin}.}
    \label{fig:supp_Bscan_with_fit}
\end{figure}

Figure \ref{fig:supp_Bscan_with_fit} shows a fit of the experimental magnetic field spectroscopy data with Eq.~\ref{eq:fr_of_Bin}. From the fit, we extract an estimate of $B_\mathrm{crit}^\parallel = \qty{66\pm 1}{mT}$ and $B_{\Phi_0} = \qty{102\pm 2}{mT}$. These values deviate from the ones predicted from the calculations presented in this section. We should first note that there is a large negative cross-correlation of $C_{1,2} \approx -0.98$ in our fitting routine between these two fitting parameters, which hinders a reliable independent extraction of both parameters at the same time; to mitigate this problem, it would be helpful to collect spectroscopy data at magnetic fields much closer to $B_\mathrm{crit}^\parallel$ or to $B_{\Phi_0}$, where the functional dependence of $\fr(\Bpar)$ allows to better fit the two parameters independently. Unfortunately, for the devices presented in this work the frequency of the resonators falls outside of our measurement bandwidth for $\Bpar > \qty{60}{mT}$. In addition to this, experimental data can deviate from the theoretical prediction presented above in the presence of two films of different thicknesses and for very thin films\cite{krause_magnetic_2022}, so that Eqs.~\ref{eq:B_crit}, and~\ref{eq:lambda_eff} remain true only qualitatively. Lastly, small deviations of the thicknesses of the Al leads from the values reported here are possible due to the fabrication process. However, estimating $B_\mathrm{crit}^\parallel$ and $B_{\Phi_0}$ with a higher precision is beyond the scope of this work.

\section{Strong coupling of JJ array resonators with TLFs}
\label{supp:strong_hybridization}
\begin{figure}[h!]
    \centering
    \includegraphics{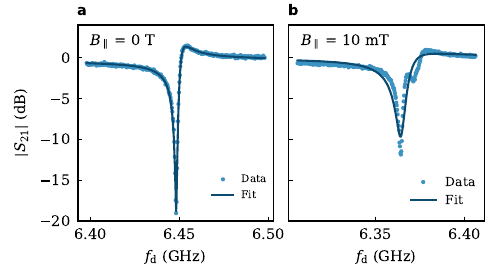}
    \caption{Spectroscopy of the JJ resonator on the intrinsic Si substrate indicated by the black arrow in Fig.~\ref{fig:magnetic_field}b. \textbf{\textsf{(a)}} is taken at zero external magnetic field, while \textbf{\textsf{(b)}} at $\Bpar = \qty{10}{mT}$. Circular blue points represent the measured data, while the continuous dark line represents a fit to Eq.~\ref{eq:S21_linear}. In \textsf{(a)}, only the resonator's resonance dip is visible. In \textsf{(b)} the resonator's frequency, lowered by the external magnetic field, comes in resonance with a TLF residing in the oxide tunneling barrier of the JJs, and couples with it, resulting in a double and less deep resonance dip. As a consequence, the fit extracts a lower internal quality factor, as indicated in Fig.~\ref{fig:magnetic_field}b.}
    \label{fig:splitting_TLF}
\end{figure}
The oxide tunneling barrier of JJs can host defects that behave as TLFs \cite{constantin_microscopic_2007, van_harlingen_decoherence_2004, de_palma_strong_2024}, whose frequency can happen to fall in the GHz range. If the frequency of a JJ array resonator comes in resonance with the frequency of a TLF, due to the strong electric field in the resonator the two systems can couple and in the resonator spectroscopy it is possible to observe a splitting of the resonator's frequency. Figure~\ref{fig:splitting_TLF} shows an example of such phenomenon for the resonators on the intrinsic Si substrate indicated by the black arrow in Fig.~\ref{fig:magnetic_field}b. In Fig.~\ref{fig:splitting_TLF}a, at $\Bpar = \qty{0}{T}$, one single resonance is visible as the resonator does not hybridize with TLFs at this frequency. However, as we apply $\Bpar = \qty{10}{mT}$ and lower the resonator's frequency, the latter comes in resonance with a TLF and observe a double resonance as shown in Fig.~\ref{fig:splitting_TLF}b. 
%Since TLFs usually have higher decoherence rates than the resonator, the coupling between them also increases the losses of the resonator, as can be observed from the smaller resonance dip in Fig.~\ref{fig:splitting_TLF}b.

%%%%%%%%%%%%%%%%%%%%%%%%%%%%%%%%%%%%%%%%%%%%%%%%%%%%%%%%%%%%%%%%%%%%%%%%%%%%%%%%%%%%%%%%%%%%%%

\nocite{*}
\bibliography{aipsamp}% Produces the bibliography via BibTeX.

@PREAMBLE{
 "\providecommand{\noopsort}[1]{}" 
 # "\providecommand{\singleletter}[1]{#1}%" 
}

@article{hendrickx_fast_2020,
    title = {Fast two-qubit logic with holes in germanium},
    volume = {577},
    copyright = {2020 The Author(s), under exclusive licence to Springer Nature Limited},
    issn = {1476-4687},
    url = {https://www.nature.com/articles/s41586-019-1919-3},
    doi = {10.1038/s41586-019-1919-3},
    
    number = {7791},
    urldate = {2024-03-11},
    journal = {Nature},
    author = {Hendrickx, N. W. and Franke, D. P. and Sammak, A. and Scappucci, G. and Veldhorst, M.},
    month = jan,
    year = {2020},
    note = {Publisher: Nature Publishing Group},
    pages = {487--491},
}

@article{hendrickx_single-hole_2020,
    title = {A single-hole spin qubit},
    volume = {11},
    copyright = {2020 The Author(s)},
    issn = {2041-1723},
    url = {https://www.nature.com/articles/s41467-020-17211-7},
    doi = {10.1038/s41467-020-17211-7},
    
    number = {1},
    urldate = {2023-10-17},
    journal = {Nature Communications},
    author = {Hendrickx, N. W. and Lawrie, W. I. L. and Petit, L. and Sammak, A. and Scappucci, G. and Veldhorst, M.},
    month = jul,
    year = {2020},
    note = {Number: 1
Publisher: Nature Publishing Group},
    pages = {3478},
}

@article{hendrickx_four-qubit_2021,
    title = {A four-qubit germanium quantum processor},
    volume = {591},
    copyright = {2021 The Author(s), under exclusive licence to Springer Nature Limited},
    issn = {1476-4687},
    url = {https://www.nature.com/articles/s41586-021-03332-6},
    doi = {10.1038/s41586-021-03332-6},
    
    number = {7851},
    urldate = {2023-10-04},
    journal = {Nature},
    author = {Hendrickx, Nico W. and Lawrie, William I. L. and Russ, Maximilian and van Riggelen, Floor and de Snoo, Sander L. and Schouten, Raymond N. and Sammak, Amir and Scappucci, Giordano and Veldhorst, Menno},
    month = mar,
    year = {2021},
    note = {Number: 7851
Publisher: Nature Publishing Group},
    pages = {580--585},
}

@article{tosato_hard_2023,
    title = {Hard superconducting gap in germanium},
    volume = {4},
    copyright = {2023 The Author(s)},
    issn = {2662-4443},
    url = {https://www.nature.com/articles/s43246-023-00351-w},
    doi = {10.1038/s43246-023-00351-w},
    
    number = {1},
    urldate = {2024-03-27},
    journal = {Communications Materials},
    author = {Tosato, Alberto and Levajac, Vukan and Wang, Ji-Yin and Boor, Casper J. and Borsoi, Francesco and Botifoll, Marc and Borja, Carla N. and Martí-Sánchez, Sara and Arbiol, Jordi and Sammak, Amir and Veldhorst, Menno and Scappucci, Giordano},
    month = apr,
    year = {2023},
    note = {Publisher: Nature Publishing Group},
    pages = {1--9},
}

@article{sammak_shallow_2019,
    title = {Shallow and {Undoped} {Germanium} {Quantum} {Wells}: {A} {Playground} for {Spin} and {Hybrid} {Quantum} {Technology}},
    volume = {29},
    issn = {1616-3028},
    shorttitle = {Shallow and {Undoped} {Germanium} {Quantum} {Wells}},
    url = {https://onlinelibrary.wiley.com/doi/abs/10.1002/adfm.201807613},
    doi = {10.1002/adfm.201807613},
    
    number = {14},
    urldate = {2024-03-08},
    journal = {Advanced Functional Materials},
    author = {Sammak, Amir and Sabbagh, Diego and Hendrickx, Nico W. and Lodari, Mario and Paquelet Wuetz, Brian and Tosato, Alberto and Yeoh, LaReine and Bollani, Monica and Virgilio, Michele and Schubert, Markus Andreas and Zaumseil, Peter and Capellini, Giovanni and Veldhorst, Menno and Scappucci, Giordano},
    year = {2019},
    note = {\_eprint: https://onlinelibrary.wiley.com/doi/pdf/10.1002/adfm.201807613},
    pages = {1807613},
}

@article{itoh_isotope_2014,
    title = {Isotope engineering of silicon and diamond for quantum computing and sensing applications},
    volume = {4},
    issn = {2159-6859, 2159-6867},
    url = {http://link.springer.com/10.1557/mrc.2014.32},
    doi = {10.1557/mrc.2014.32},
    
    number = {4},
    urldate = {2025-10-10},
    journal = {MRS Communications},
    author = {Itoh, Kohei M. and Watanabe, Hideyuki},
    month = dec,
    year = {2014},
    pages = {143--157},
}

@article{blais_circuit_2021,
    title = {Circuit quantum electrodynamics},
    volume = {93},
    url = {https://link.aps.org/doi/10.1103/RevModPhys.93.025005},
    doi = {10.1103/RevModPhys.93.025005},
    number = {2},
    urldate = {2024-04-09},
    journal = {Reviews of Modern Physics},
    author = {Blais, Alexandre and Grimsmo, Arne L. and Girvin, S. M. and Wallraff, Andreas},
    month = may,
    year = {2021},
    note = {Publisher: American Physical Society},
    pages = {025005},
}

@article{harvey-collard_coherent_2022,
    title = {Coherent {Spin}-{Spin} {Coupling} {Mediated} by {Virtual} {Microwave} {Photons}},
    volume = {12},
    issn = {2160-3308},
    url = {https://link.aps.org/doi/10.1103/PhysRevX.12.021026},
    doi = {10.1103/PhysRevX.12.021026},
    
    number = {2},
    urldate = {2025-08-04},
    journal = {Physical Review X},
    author = {Harvey-Collard, Patrick and Dijkema, Jurgen and Zheng, Guoji and Sammak, Amir and Scappucci, Giordano and Vandersypen, Lieven M. K.},
    month = may,
    year = {2022},
    pages = {021026},
}

@article{dijkema_cavity-mediated_2025,
    title = {Cavity-mediated {iSWAP} oscillations between distant spins},
    volume = {21},
    copyright = {https://creativecommons.org/licenses/by/4.0},
    issn = {1745-2473, 1745-2481},
    url = {https://www.nature.com/articles/s41567-024-02694-8},
    doi = {10.1038/s41567-024-02694-8},
    
    number = {1},
    urldate = {2025-07-08},
    journal = {Nature Physics},
    author = {Dijkema, Jurgen and Xue, Xiao and Harvey-Collard, Patrick and Rimbach-Russ, Maximilian and De Snoo, Sander L. and Zheng, Guoji and Sammak, Amir and Scappucci, Giordano and Vandersypen, Lieven M. K.},
    month = jan,
    year = {2025},
    note = {Publisher: Springer Science and Business Media LLC},
    pages = {168--174},
}

@article{bargerbos_spectroscopy_2023,
    title = {Spectroscopy of {Spin}-{Split} {Andreev} {Levels} in a {Quantum} {Dot} with {Superconducting} {Leads}},
    volume = {131},
    issn = {0031-9007, 1079-7114},
    url = {https://link.aps.org/doi/10.1103/PhysRevLett.131.097001},
    doi = {10.1103/PhysRevLett.131.097001},
    
    number = {9},
    urldate = {2025-10-10},
    journal = {Physical Review Letters},
    author = {Bargerbos, Arno and Pita-Vidal, Marta and Žitko, Rok and Splitthoff, Lukas J. and Grünhaupt, Lukas and Wesdorp, Jaap J. and Liu, Yu and Kouwenhoven, Leo P. and Aguado, Ramón and Andersen, Christian Kraglund and Kou, Angela and Van Heck, Bernard},
    month = aug,
    year = {2023},
    pages = {097001},
}

@article{pita-vidal_direct_2023,
    title = {Direct manipulation of a superconducting spin qubit strongly coupled to a transmon qubit},
    volume = {19},
    issn = {1745-2473, 1745-2481},
    url = {https://www.nature.com/articles/s41567-023-02071-x},
    doi = {10.1038/s41567-023-02071-x},
    
    number = {8},
    urldate = {2025-10-10},
    journal = {Nature Physics},
    author = {Pita-Vidal, Marta and Bargerbos, Arno and Žitko, Rok and Splitthoff, Lukas J. and Grünhaupt, Lukas and Wesdorp, Jaap J. and Liu, Yu and Kouwenhoven, Leo P. and Aguado, Ramón and Van Heck, Bernard and Kou, Angela and Andersen, Christian Kraglund},
    month = aug,
    year = {2023},
    pages = {1110--1115},
}

@article{wesdorp_microwave_2024,
    title = {Microwave spectroscopy of interacting {Andreev} spins},
    volume = {109},
    issn = {2469-9950, 2469-9969},
    url = {https://link.aps.org/doi/10.1103/PhysRevB.109.045302},
    doi = {10.1103/PhysRevB.109.045302},
    
    number = {4},
    urldate = {2025-10-10},
    journal = {Physical Review B},
    author = {Wesdorp, J. J. and Matute-Cañadas, F. J. and Vaartjes, A. and Grünhaupt, L. and Laeven, T. and Roelofs, S. and Splitthoff, L. J. and Pita-Vidal, M. and Bargerbos, A. and Van Woerkom, D. J. and Krogstrup, P. and Kouwenhoven, L. P. and Andersen, C. K. and Yeyati, A. Levy and Van Heck, B. and De Lange, G.},
    month = jan,
    year = {2024},
    pages = {045302},
}

@article{de_palma_strong_2024,
    title = {Strong hole-photon coupling in planar {Ge} for probing charge degree and strongly correlated states},
    volume = {15},
    issn = {2041-1723},
    url = {https://www.nature.com/articles/s41467-024-54520-7},
    doi = {10.1038/s41467-024-54520-7},
    
    number = {1},
    urldate = {2025-07-04},
    journal = {Nature Communications},
    author = {De Palma, Franco and Oppliger, Fabian and Jang, Wonjin and Bosco, Stefano and Janík, Marián and Calcaterra, Stefano and Katsaros, Georgios and Isella, Giovanni and Loss, Daniel and Scarlino, Pasquale},
    month = nov,
    year = {2024},
    pages = {10177},
}

@article{janik_strong_2025,
    title = {Strong charge-photon coupling in planar germanium enabled by granular aluminium superinductors},
    volume = {16},
    issn = {2041-1723},
    url = {https://www.nature.com/articles/s41467-025-57252-4},
    doi = {10.1038/s41467-025-57252-4},
    
    number = {1},
    urldate = {2025-07-04},
    journal = {Nature Communications},
    author = {Janík, Marián and Roux, Kevin and Borja-Espinosa, Carla and Sagi, Oliver and Baghdadi, Abdulhamid and Adletzberger, Thomas and Calcaterra, Stefano and Botifoll, Marc and Garzón Manjón, Alba and Arbiol, Jordi and Chrastina, Daniel and Isella, Giovanni and Pop, Ioan M. and Katsaros, Georgios},
    month = mar,
    year = {2025},
    pages = {2103},
}

@article{nigro_demonstration_2024,
    title = {Demonstration of {Microwave} {Resonators} and {Double} {Quantum} {Dots} on {Optimized} {Reverse}-{Graded} {Ge}/{SiGe} {Heterostructures}},
    volume = {6},
    copyright = {https://creativecommons.org/licenses/by/4.0/},
    issn = {2637-6113, 2637-6113},
    url = {https://pubs.acs.org/doi/10.1021/acsaelm.4c00654},
    doi = {10.1021/acsaelm.4c00654},
    
    number = {7},
    urldate = {2025-05-05},
    journal = {ACS Applied Electronic Materials},
    author = {Nigro, Arianna and Jutzi, Eric and Oppliger, Fabian and De Palma, Franco and Olsen, Christian and Ruiz-Caridad, Alicia and Gadea, Gerard and Scarlino, Pasquale and Zardo, Ilaria and Hofmann, Andrea},
    month = jul,
    year = {2024},
    pages = {5094--5100},
}

@article{krause_magnetic_2022,
    title = {Magnetic {Field} {Resilience} of {Three}-{Dimensional} {Transmons} with {Thin}-{Film} {Al}/{AlOx}/{Al} {Josephson} {Junctions} {Approaching} 1 {T}},
    volume = {17},
    url = {https://link.aps.org/doi/10.1103/PhysRevApplied.17.034032},
    doi = {10.1103/PhysRevApplied.17.034032},
    number = {3},
    urldate = {2024-04-11},
    journal = {Physical Review Applied},
    author = {Krause, J. and Dickel, C. and Vaal, E. and Vielmetter, M. and Feng, J. and Bounds, R. and Catelani, G. and Fink, J. M. and Ando, Yoichi},
    month = mar,
    year = {2022},
    note = {Publisher: American Physical Society},
    pages = {034032},
}

@article{zhang_universal_2025,
    title = {Universal control of four singlet–triplet qubits},
    volume = {20},
    issn = {1748-3387, 1748-3395},
    url = {https://www.nature.com/articles/s41565-024-01817-9},
    doi = {10.1038/s41565-024-01817-9},
    
    number = {2},
    urldate = {2025-10-01},
    journal = {Nature Nanotechnology},
    author = {Zhang, Xin and Morozova, Elizaveta and Rimbach-Russ, Maximilian and Jirovec, Daniel and Hsiao, Tzu-Kan and Fariña, Pablo Cova and Wang, Chien-An and Oosterhout, Stefan D. and Sammak, Amir and Scappucci, Giordano and Veldhorst, Menno and Vandersypen, Lieven M. K.},
    month = feb,
    year = {2025},
    pages = {209--215},
}

@article{kuzmin_tuning_2023,
    title = {Tuning the inductance of {Josephson} junction arrays without {SQUIDs}},
    volume = {123},
    issn = {0003-6951, 1077-3118},
    url = {https://pubs.aip.org/apl/article/123/18/182602/2919916/Tuning-the-inductance-of-Josephson-junction-arrays},
    doi = {10.1063/5.0171047},
    
    number = {18},
    urldate = {2024-10-28},
    journal = {Applied Physics Letters},
    author = {Kuzmin, R. and Mehta, N. and Grabon, N. and Manucharyan, V. E.},
    month = oct,
    year = {2023},
    pages = {182602},
}

@article{kuzmin_quantum_2019,
    title = {Quantum electrodynamics of a superconductor–insulator phase transition},
    volume = {15},
    issn = {1745-2473, 1745-2481},
    url = {https://www.nature.com/articles/s41567-019-0553-1},
    doi = {10.1038/s41567-019-0553-1},
    
    number = {9},
    urldate = {2024-10-28},
    journal = {Nature Physics},
    author = {Kuzmin, R. and Mencia, R. and Grabon, N. and Mehta, N. and Lin, Y.-H. and Manucharyan, V. E.},
    month = sep,
    year = {2019},
    pages = {930--934},
}

@article{pita-vidal_strong_2024,
    title = {Strong tunable coupling between two distant superconducting spin qubits},
    volume = {20},
    issn = {1745-2473, 1745-2481},
    url = {https://www.nature.com/articles/s41567-024-02497-x},
    doi = {10.1038/s41567-024-02497-x},
    
    number = {7},
    urldate = {2025-10-10},
    journal = {Nature Physics},
    author = {Pita-Vidal, Marta and Wesdorp, Jaap J. and Splitthoff, Lukas J. and Bargerbos, Arno and Liu, Yu and Kouwenhoven, Leo P. and Andersen, Christian Kraglund},
    month = jul,
    year = {2024},
    pages = {1158--1163},
}

@article{lodari_lightly_2022,
    title = {Lightly strained germanium quantum wells with hole mobility exceeding one million},
    volume = {120},
    issn = {0003-6951},
    url = {https://doi.org/10.1063/5.0083161},
    doi = {10.1063/5.0083161},
    number = {12},
    urldate = {2025-10-17},
    journal = {Applied Physics Letters},
    author = {Lodari, M. and Kong, O. and Rendell, M. and Tosato, A. and Sammak, A. and Veldhorst, M. and Hamilton, A. R. and Scappucci, G.},
    month = mar,
    year = {2022},
    pages = {122104},
}

@article{moutanabbir_nuclear_2024,
    title = {Nuclear {Spin}-{Depleted}, {Isotopically} {Enriched} {70Ge}/{28Si70Ge} {Quantum} {Wells}},
    volume = {36},
    copyright = {© 2023 The Authors. Advanced Materials published by Wiley-VCH GmbH},
    issn = {1521-4095},
    url = {https://onlinelibrary.wiley.com/doi/abs/10.1002/adma.202305703},
    doi = {10.1002/adma.202305703},
    
    number = {8},
    urldate = {2025-10-17},
    journal = {Advanced Materials},
    author = {Moutanabbir, Oussama and Assali, Simone and Attiaoui, Anis and Daligou, Gérard and Daoust, Patrick and Vecchio, Patrick Del and Koelling, Sebastian and Luo, Lu and Rotaru, Nicolas},
    year = {2024},
    note = {\_eprint: https://advanced.onlinelibrary.wiley.com/doi/pdf/10.1002/adma.202305703},
    pages = {2305703},
}

@article{fischer_spin_2008,
    title = {Spin decoherence of a heavy hole coupled to nuclear spins in a quantum dot},
    volume = {78},
    doi = {10.1103/PhysRevB.78.155329},
    number = {15},
    journal = {Physical Review B},
    author = {Fischer, Jan},
    year = {2008},
}

@article{terrazos_theory_2021,
    title = {Theory of hole-spin qubits in strained germanium quantum dots},
    volume = {103},
    url = {https://link.aps.org/doi/10.1103/PhysRevB.103.125201},
    doi = {10.1103/PhysRevB.103.125201},
    number = {12},
    urldate = {2025-10-17},
    journal = {Physical Review B},
    author = {Terrazos, L. A. and Marcellina, E. and Wang, Zhanning and Coppersmith, S. N. and Friesen, Mark and Hamilton, A. R. and Hu, Xuedong and Koiller, Belita and Saraiva, A. L. and Culcer, Dimitrie and Capaz, Rodrigo B.},
    month = mar,
    year = {2021},
    note = {Publisher: American Physical Society},
    pages = {125201},
}

@article{hendrickx_sweet-spot_2024,
    title = {Sweet-spot operation of a germanium hole spin qubit with highly anisotropic noise sensitivity},
    volume = {23},
    copyright = {2024 The Author(s)},
    issn = {1476-4660},
    url = {https://www.nature.com/articles/s41563-024-01857-5},
    doi = {10.1038/s41563-024-01857-5},
    
    number = {7},
    urldate = {2025-10-17},
    journal = {Nature Materials},
    author = {Hendrickx, N. W. and Massai, L. and Mergenthaler, M. and Schupp, F. J. and Paredes, S. and Bedell, S. W. and Salis, G. and Fuhrer, A.},
    month = jul,
    year = {2024},
    note = {Publisher: Nature Publishing Group},
    pages = {920--927},
}

@article{scappucci_germanium_2021,
    title = {The germanium quantum information route},
    volume = {6},
    copyright = {2020 Springer Nature Limited},
    issn = {2058-8437},
    url = {https://www.nature.com/articles/s41578-020-00262-z},
    doi = {10.1038/s41578-020-00262-z},
    
    number = {10},
    urldate = {2025-10-17},
    journal = {Nature Reviews Materials},
    author = {Scappucci, Giordano and Kloeffel, Christoph and Zwanenburg, Floris A. and Loss, Daniel and Myronov, Maksym and Zhang, Jian-Jun and De Franceschi, Silvano and Katsaros, Georgios and Veldhorst, Menno},
    month = oct,
    year = {2021},
    note = {Publisher: Nature Publishing Group},
    pages = {926--943},
}

@article{dvir_realization_2023,
    title = {Realization of a minimal {Kitaev} chain in coupled quantum dots},
    volume = {614},
    copyright = {2023 The Author(s), under exclusive licence to Springer Nature Limited},
    issn = {1476-4687},
    url = {https://www.nature.com/articles/s41586-022-05585-1},
    doi = {10.1038/s41586-022-05585-1},
    
    number = {7948},
    urldate = {2025-10-17},
    journal = {Nature},
    author = {Dvir, Tom and Wang, Guanzhong and van Loo, Nick and Liu, Chun-Xiao and Mazur, Grzegorz P. and Bordin, Alberto and ten Haaf, Sebastiaan L. D. and Wang, Ji-Yin and van Driel, David and Zatelli, Francesco and Li, Xiang and Malinowski, Filip K. and Gazibegovic, Sasa and Badawy, Ghada and Bakkers, Erik P. A. M. and Wimmer, Michael and Kouwenhoven, Leo P.},
    month = feb,
    year = {2023},
    note = {Publisher: Nature Publishing Group},
    pages = {445--450},
}

@article{hinderling_direct_2024,
    title = {Direct {Microwave} {Spectroscopy} of {Andreev} {Bound} {States} in {Planar} \${\textbackslash}mathrm\{{Ge}\}\$ {Josephson} {Junctions}},
    volume = {5},
    url = {https://link.aps.org/doi/10.1103/PRXQuantum.5.030357},
    doi = {10.1103/PRXQuantum.5.030357},
    number = {3},
    urldate = {2025-10-17},
    journal = {PRX Quantum},
    author = {Hinderling, M. and ten Kate, S. C. and Coraiola, M. and Haxell, D.Z. and Stiefel, M. and Mergenthaler, M. and Paredes, S. and Bedell, S.W. and Sabonis, D. and Nichele, F.},
    month = sep,
    year = {2024},
    note = {Publisher: American Physical Society},
    pages = {030357},
}

@article{zheng_rapid_2019,
    title = {Rapid gate-based spin read-out in silicon using an on-chip resonator},
    volume = {14},
    copyright = {2019 The Author(s), under exclusive licence to Springer Nature Limited},
    issn = {1748-3395},
    url = {https://www.nature.com/articles/s41565-019-0488-9},
    doi = {10.1038/s41565-019-0488-9},
    
    number = {8},
    urldate = {2025-10-17},
    journal = {Nature Nanotechnology},
    author = {Zheng, Guoji and Samkharadze, Nodar and Noordam, Marc L. and Kalhor, Nima and Brousse, Delphine and Sammak, Amir and Scappucci, Giordano and Vandersypen, Lieven M. K.},
    month = aug,
    year = {2019},
    note = {Publisher: Nature Publishing Group},
    pages = {742--746},
}

@misc{oppliger_high-efficiency_2025,
    title = {High-{Efficiency} {Tunable} {Microwave} {Photon} {Detector} {Based} on a {Semiconductor} {Double} {Quantum} {Dot} {Coupled} to a {Superconducting} {High}-{Impedance} {Cavity}},
    url = {http://arxiv.org/abs/2506.19828},
    doi = {10.48550/arXiv.2506.19828},
    urldate = {2025-10-17},
    publisher = {arXiv},
    author = {Oppliger, Fabian and Jang, Wonjin and Tarascio, Aldo and Palma, Franco De and Reichl, Christian and Wegscheider, Werner and Maisi, Ville F. and Zumbühl, Dominik and Scarlino, Pasquale},
    month = jun,
    year = {2025},
    note = {arXiv:2506.19828 [quant-ph]},
}

@article{liu_semiconductor_2015,
    title = {Semiconductor double quantum dot micromaser},
    volume = {347},
    url = {https://www.science.org/doi/10.1126/science.aaa2501},
    doi = {10.1126/science.aaa2501},
    number = {6219},
    urldate = {2025-10-17},
    journal = {Science},
    author = {Liu, Y.-Y. and Stehlik, J. and Eichler, C. and Gullans, M. J. and Taylor, J. M. and Petta, J. R.},
    month = jan,
    year = {2015},
    note = {Publisher: American Association for the Advancement of Science},
    pages = {285--287},
}

@article{yu_strong_2023,
    title = {Strong coupling between a photon and a hole spin in silicon},
    volume = {18},
    copyright = {2023 The Author(s), under exclusive licence to Springer Nature Limited},
    issn = {1748-3395},
    url = {https://www.nature.com/articles/s41565-023-01332-3},
    doi = {10.1038/s41565-023-01332-3},
    
    number = {7},
    urldate = {2025-10-17},
    journal = {Nature Nanotechnology},
    author = {Yu, Cécile X. and Zihlmann, Simon and Abadillo-Uriel, José C. and Michal, Vincent P. and Rambal, Nils and Niebojewski, Heimanu and Bedecarrats, Thomas and Vinet, Maud and Dumur, \'Etienne and Filippone, Michele and Bertrand, Benoit and De Franceschi, Silvano and Niquet, Yann-Michel and Maurand, Romain},
    month = jul,
    year = {2023},
    note = {Publisher: Nature Publishing Group},
    pages = {741--746},
}

@article{burkard_superconductorsemiconductor_2020,
    title = {Superconductor–semiconductor hybrid-circuit quantum electrodynamics},
    volume = {2},
    copyright = {2020 Springer Nature Limited},
    issn = {2522-5820},
    url = {https://www.nature.com/articles/s42254-019-0135-2},
    doi = {10.1038/s42254-019-0135-2},
    
    number = {3},
    urldate = {2025-10-17},
    journal = {Nature Reviews Physics},
    author = {Burkard, Guido and Gullans, Michael J. and Mi, Xiao and Petta, Jason R.},
    month = mar,
    year = {2020},
    note = {Publisher: Nature Publishing Group},
    pages = {129--140},
}

@article{frunzio_fabrication_2005,
    title = {Fabrication and characterization of superconducting circuit {QED} devices for quantum computation},
    volume = {15},
    issn = {1558-2515},
    url = {https://ieeexplore.ieee.org/document/1439774/},
    doi = {10.1109/TASC.2005.850084},
    number = {2},
    urldate = {2025-10-17},
    journal = {IEEE Transactions on Applied Superconductivity},
    author = {Frunzio, L. and Wallraff, A. and Schuster, D. and Majer, J. and Schoelkopf, R.},
    month = jun,
    year = {2005},
    pages = {860--863},
}

@article{peyruchat_landauzener_2025,
    title = {Landau–{Zener} without a qubit: multiphoton sidebands interaction and signatures of dissipative quantum chaos},
    volume = {11},
    copyright = {2025 The Author(s)},
    issn = {2056-6387},
    shorttitle = {Landau–{Zener} without a qubit},
    url = {https://www.nature.com/articles/s41534-025-00984-4},
    doi = {10.1038/s41534-025-00984-4},
    
    number = {1},
    urldate = {2025-10-17},
    journal = {npj Quantum Information},
    author = {Peyruchat, Léo and Minganti, Fabrizio and Scigliuzzo, Marco and Ferrari, Filippo and Jouanny, Vincent and Nori, Franco and Savona, Vincenzo and Scarlino, Pasquale},
    month = apr,
    year = {2025},
    note = {Publisher: Nature Publishing Group},
    pages = {62},
}

@article{sagi_gate_2024,
    title = {A gate tunable transmon qubit in planar {Ge}},
    volume = {15},
    copyright = {2024 The Author(s)},
    issn = {2041-1723},
    url = {https://www.nature.com/articles/s41467-024-50763-6},
    doi = {10.1038/s41467-024-50763-6},
    
    number = {1},
    urldate = {2025-10-17},
    journal = {Nature Communications},
    author = {Sagi, Oliver and Crippa, Alessandro and Valentini, Marco and Janik, Marian and Baghumyan, Levon and Fabris, Giorgio and Kapoor, Lucky and Hassani, Farid and Fink, Johannes and Calcaterra, Stefano and Chrastina, Daniel and Isella, Giovanni and Katsaros, Georgios},
    month = jul,
    year = {2024},
    note = {Publisher: Nature Publishing Group},
    pages = {6400},
}

@article{lee_strained_2004,
    title = {Strained {Si}, {SiGe}, and {Ge} channels for high-mobility metal-oxide-semiconductor field-effect transistors},
    volume = {97},
    issn = {0021-8979},
    url = {https://doi.org/10.1063/1.1819976},
    doi = {10.1063/1.1819976},
    number = {1},
    urldate = {2025-10-17},
    journal = {Journal of Applied Physics},
    author = {Lee, Minjoo L. and Fitzgerald, Eugene A. and Bulsara, Mayank T. and Currie, Matthew T. and Lochtefeld, Anthony},
    month = dec,
    year = {2004},
    pages = {011101},
}

@article{frasca_nbn_2023,
    title = {{NbN} films with high kinetic inductance for high-quality compact superconducting resonators},
    volume = {20},
    url = {https://link.aps.org/doi/10.1103/PhysRevApplied.20.044021},
    doi = {10.1103/PhysRevApplied.20.044021},
    number = {4},
    urldate = {2025-10-17},
    journal = {Physical Review Applied},
    author = {Frasca, S. and Arabadzhiev, I.N. and de Puechredon, S.Y. Bros and Oppliger, F. and Jouanny, V. and Musio, R. and Scigliuzzo, M. and Minganti, F. and Scarlino, P. and Charbon, E.},
    month = oct,
    year = {2023},
    note = {Publisher: American Physical Society},
    pages = {044021},
}

@article{mcrae_materials_2020,
    title = {Materials loss measurements using superconducting microwave resonators},
    volume = {91},
    issn = {0034-6748},
    url = {https://doi.org/10.1063/5.0017378},
    doi = {10.1063/5.0017378},
    number = {9},
    urldate = {2025-10-17},
    journal = {Review of Scientific Instruments},
    author = {McRae, C. R. H. and Wang, H. and Gao, J. and Vissers, M. R. and Brecht, T. and Dunsworth, A. and Pappas, D. P. and Mutus, J.},
    month = sep,
    year = {2020},
    pages = {091101},
}

@article{dolan_offset_1977,
    title = {Offset masks for lift‐off photoprocessing},
    volume = {31},
    issn = {0003-6951},
    url = {https://doi.org/10.1063/1.89690},
    doi = {10.1063/1.89690},
    number = {5},
    urldate = {2025-10-17},
    journal = {Applied Physics Letters},
    author = {Dolan, G. J.},
    month = sep,
    year = {1977},
    pages = {337--339},
}

@article{van_harlingen_decoherence_2004,
    title = {Decoherence in {Josephson}-junction qubits due to critical-current fluctuations},
    volume = {70},
    copyright = {http://link.aps.org/licenses/aps-default-license},
    issn = {1098-0121, 1550-235X},
    url = {https://link.aps.org/doi/10.1103/PhysRevB.70.064517},
    doi = {10.1103/PhysRevB.70.064517},
    
    number = {6},
    urldate = {2025-10-17},
    journal = {Physical Review B},
    author = {Van Harlingen, D. J. and Robertson, T. L. and Plourde, B. L. T. and Reichardt, P. A. and Crane, T. A. and Clarke, John},
    month = aug,
    year = {2004},
    pages = {064517},
}

@article{constantin_microscopic_2007,
    title = {Microscopic {Model} of {Critical} {Current} {Noise} in {Josephson} {Junctions}},
    volume = {99},
    copyright = {http://link.aps.org/licenses/aps-default-license},
    issn = {0031-9007, 1079-7114},
    url = {https://link.aps.org/doi/10.1103/PhysRevLett.99.207001},
    doi = {10.1103/PhysRevLett.99.207001},
    
    number = {20},
    urldate = {2025-10-17},
    journal = {Physical Review Letters},
    author = {Constantin, Magdalena and Yu, Clare C.},
    month = nov,
    year = {2007},
    pages = {207001},
}

@article{eichler_controlling_2014,
    title = {Controlling the dynamic range of a {Josephson} parametric amplifier},
    volume = {1},
    issn = {2196-0763},
    url = {http://www.epjquantumtechnology.com/content/1/1/2},
    doi = {10.1140/epjqt2},
    
    number = {1},
    urldate = {2025-07-04},
    journal = {EPJ Quantum Technology},
    author = {Eichler, Christopher and Wallraff, Andreas},
    month = dec,
    year = {2014},
    pages = {2},
}

@book{tinkham_introduction_1996,
    address = {New York},
    title = {Introduction to {Superconductivity}},
    publisher = {McGraw Hill},
    author = {Tinkham, Michael},
    year = {1996},
}

@phdthesis{winkel_superconducting_nodate,
    title = {Superconducting quantum circuits for hybrid architectures},
    year = {2020},
    school = {Karlsruher Instituts für Technologie (KIT)},
    author = {Winkel, Patrick},
}

@article{chen_scattering_2022,
    title = {Scattering coefficients of superconducting microwave resonators. {II}. {System}-bath approach},
    volume = {106},
    issn = {2469-9950, 2469-9969},
    url = {https://link.aps.org/doi/10.1103/PhysRevB.106.214506},
    doi = {10.1103/PhysRevB.106.214506},
    
    number = {21},
    urldate = {2025-10-22},
    journal = {Physical Review B},
    author = {Chen, Qi-Ming and Partanen, Matti and Fesquet, Florian and Honasoge, Kedar E. and Kronowetter, Fabian and Nojiri, Yuki and Renger, Michael and Fedorov, Kirill G. and Marx, Achim and Deppe, Frank and Gross, Rudolf},
    month = dec,
    year = {2022},
    pages = {214506},
}

@article{gardiner_input_1985,
    title = {Input and output in damped quantum systems: {Quantum} stochastic differential equations and the master equation},
    volume = {31},
    copyright = {http://link.aps.org/licenses/aps-default-license},
    issn = {0556-2791},
    shorttitle = {Input and output in damped quantum systems},
    url = {https://link.aps.org/doi/10.1103/PhysRevA.31.3761},
    doi = {10.1103/PhysRevA.31.3761},
    
    number = {6},
    urldate = {2025-10-22},
    journal = {Physical Review A},
    author = {Gardiner, C. W. and Collett, M. J.},
    month = jun,
    year = {1985},
    pages = {3761--3774},
}

@article{probst_efficient_2015,
    title = {Efficient and robust analysis of complex scattering data under noise in microwave resonators},
    volume = {86},
    issn = {0034-6748, 1089-7623},
    url = {http://arxiv.org/abs/1410.3365},
    doi = {10.1063/1.4907935},
    number = {2},
    urldate = {2024-03-18},
    journal = {Review of Scientific Instruments},
    author = {Probst, S. and Song, F. B. and Bushev, P. A. and Ustinov, A. V. and Weides, M.},
    month = feb,
    year = {2015},
    note = {arXiv:1410.3365 [cond-mat, physics:physics]},
    pages = {024706},
}

@misc{roy_study_2025,
    title = {Study of {Magnetic} {Field} {Resilient} {High} {Impedance} {High}-{Kinetic} {Inductance} {Superconducting} {Resonators}},
    url = {http://arxiv.org/abs/2503.13321},
    doi = {10.48550/arXiv.2503.13321},
    
    urldate = {2025-05-26},
    publisher = {arXiv},
    author = {Roy, Camille and Frasca, Simone and Scarlino, Pasquale},
    month = mar,
    year = {2025},
    note = {arXiv:2503.13321 [quant-ph]},
}

@article{anferov_millimeter-wave_2020,
    title = {Millimeter-{Wave} {Four}-{Wave} {Mixing} via {Kinetic} {Inductance} for {Quantum} {Devices}},
    volume = {13},
    issn = {2331-7019},
    url = {https://link.aps.org/doi/10.1103/PhysRevApplied.13.024056},
    doi = {10.1103/PhysRevApplied.13.024056},
    
    number = {2},
    urldate = {2025-07-04},
    journal = {Physical Review Applied},
    author = {Anferov, Alexander and Suleymanzade, Aziza and Oriani, Andrew and Simon, Jonathan and Schuster, David I.},
    month = feb,
    year = {2020},
    pages = {024056},
}

@article{jirovec_singlet-triplet_2021,
    title = {A singlet-triplet hole spin qubit in planar {Ge}},
    volume = {20},
    copyright = {2021 The Author(s), under exclusive licence to Springer Nature Limited},
    issn = {1476-4660},
    url = {https://www.nature.com/articles/s41563-021-01022-2},
    doi = {10.1038/s41563-021-01022-2},
    
    number = {8},
    urldate = {2025-10-28},
    journal = {Nature Materials},
    author = {Jirovec, Daniel and Hofmann, Andrea and Ballabio, Andrea and Mutter, Philipp M. and Tavani, Giulio and Botifoll, Marc and Crippa, Alessandro and Kukucka, Josip and Sagi, Oliver and Martins, Frederico and Saez-Mollejo, Jaime and Prieto, Ivan and Borovkov, Maksim and Arbiol, Jordi and Chrastina, Daniel and Isella, Giovanni and Katsaros, Georgios},
    month = aug,
    year = {2021},
    note = {Publisher: Nature Publishing Group},
    pages = {1106--1112},
}

@article{shah_reverse_2008,
    title = {Reverse graded relaxed buffers for high {Ge} content {SiGe} virtual substrates},
    volume = {93},
    issn = {0003-6951},
    url = {https://doi.org/10.1063/1.3023068},
    doi = {10.1063/1.3023068},
    number = {19},
    urldate = {2025-10-29},
    journal = {Applied Physics Letters},
    author = {Shah, V. A. and Dobbie, A. and Myronov, M. and Fulgoni, D. J. F. and Nash, L. J. and Leadley, D. R.},
    month = nov,
    year = {2008},
    pages = {192103},
}

@article{nigro_high_2024,
    title = {High quality {Ge} layers for {Ge}/{SiGe} quantum well heterostructures using chemical vapor deposition},
    volume = {8},
    issn = {2475-9953},
    url = {https://link.aps.org/doi/10.1103/PhysRevMaterials.8.066201},
    doi = {10.1103/PhysRevMaterials.8.066201},
    
    number = {6},
    urldate = {2025-10-29},
    journal = {Physical Review Materials},
    author = {Nigro, Arianna and Jutzi, Eric and Forrer, Nicolas and Hofmann, Andrea and Gadea, Gerard and Zardo, Ilaria},
    month = jun,
    year = {2024},
    pages = {066201},
}

@article{martinis_decoherence_2005,
    title = {Decoherence in {Josephson} {Qubits} from {Dielectric} {Loss}},
    volume = {95},
    copyright = {http://link.aps.org/licenses/aps-default-license},
    issn = {0031-9007, 1079-7114},
    url = {https://link.aps.org/doi/10.1103/PhysRevLett.95.210503},
    doi = {10.1103/PhysRevLett.95.210503},
    
    number = {21},
    urldate = {2025-10-29},
    journal = {Physical Review Letters},
    author = {Martinis, John M. and Cooper, K. B. and McDermott, R. and Steffen, Matthias and Ansmann, Markus and Osborn, K. D. and Cicak, K. and Oh, Seongshik and Pappas, D. P. and Simmonds, R. W. and Yu, Clare C.},
    month = nov,
    year = {2005},
    pages = {210503},
}

@article{burnett_evidence_2014,
    title = {Evidence for interacting two-level systems from the 1/f noise of a superconducting resonator},
    volume = {5},
    copyright = {2014 Springer Nature Limited},
    issn = {2041-1723},
    url = {https://www.nature.com/articles/ncomms5119},
    doi = {10.1038/ncomms5119},
    
    number = {1},
    urldate = {2025-10-29},
    journal = {Nature Communications},
    author = {Burnett, J. and Faoro, L. and Wisby, I. and Gurtovoi, V. L. and Chernykh, A. V. and Mikhailov, G. M. and Tulin, V. A. and Shaikhaidarov, R. and Antonov, V. and Meeson, P. J. and Tzalenchuk, A. Ya and Lindström, T.},
    month = jun,
    year = {2014},
    note = {Publisher: Nature Publishing Group},
    pages = {4119},
}

@article{de_graaf_suppression_2018,
    title = {Suppression of low-frequency charge noise in superconducting resonators by surface spin desorption},
    volume = {9},
    copyright = {2018 The Author(s)},
    issn = {2041-1723},
    url = {https://www.nature.com/articles/s41467-018-03577-2},
    doi = {10.1038/s41467-018-03577-2},
    
    number = {1},
    urldate = {2025-10-29},
    journal = {Nature Communications},
    author = {de Graaf, S. E. and Faoro, L. and Burnett, J. and Adamyan, A. A. and Tzalenchuk, A. Ya and Kubatkin, S. E. and Lindström, T. and Danilov, A. V.},
    month = mar,
    year = {2018},
    note = {Publisher: Nature Publishing Group},
    pages = {1143},
}

@article{de_sousa_dangling-bond_2007,
    title = {Dangling-bond spin relaxation and magnetic $1/f$ noise from the amorphous-semiconductor/oxide interface: {Theory}},
    volume = {76},
    shorttitle = {Dangling-bond spin relaxation and magnetic $1/f$ noise from the amorphous-semiconductor/oxide interface},
    url = {https://link.aps.org/doi/10.1103/PhysRevB.76.245306},
    doi = {10.1103/PhysRevB.76.245306},
    number = {24},
    urldate = {2025-10-29},
    journal = {Physical Review B},
    author = {de Sousa, Rogerio},
    month = dec,
    year = {2007},
    note = {Publisher: American Physical Society},
    pages = {245306},
}

@article{mendes_martins_defect_2023,
    title = {Defect {Profiling} of {Oxide}-{Semiconductor} {Interfaces} {Using} {Low}-{Energy} {Muons}},
    volume = {10},
    copyright = {© 2023 The Authors. Advanced Materials Interfaces published by Wiley-VCH GmbH},
    issn = {2196-7350},
    url = {https://onlinelibrary.wiley.com/doi/abs/10.1002/admi.202300209},
    doi = {10.1002/admi.202300209},
    
    number = {21},
    urldate = {2025-10-29},
    journal = {Advanced Materials Interfaces},
    author = {Mendes Martins, Maria and Kumar, Piyush and Woerle, Judith and Ni, Xiaojie and Grossner, Ulrike and Prokscha, Thomas},
    year = {2023},
    note = {\_eprint: https://advanced.onlinelibrary.wiley.com/doi/pdf/10.1002/admi.202300209},
    pages = {2300209},
}

@article{scigliuzzo_phononic_2020,
    title = {Phononic loss in superconducting resonators on piezoelectric substrates},
    volume = {22},
    issn = {1367-2630},
    url = {https://iopscience.iop.org/article/10.1088/1367-2630/ab8044},
    doi = {10.1088/1367-2630/ab8044},
    
    number = {5},
    urldate = {2025-10-31},
    journal = {New Journal of Physics},
    author = {Scigliuzzo, Marco and Bruhat, Laure E and Bengtsson, Andreas and Burnett, Jonathan J and Roudsari, Anita Fadavi and Delsing, Per},
    month = may,
    year = {2020},
    pages = {053027},
}

@article{burnett_analysis_2016,
    title = {Analysis of high quality superconducting resonators: consequences for {TLS} properties in amorphous oxides},
    volume = {29},
    issn = {0953-2048, 1361-6668},
    shorttitle = {Analysis of high quality superconducting resonators},
    url = {https://iopscience.iop.org/article/10.1088/0953-2048/29/4/044008},
    doi = {10.1088/0953-2048/29/4/044008},
    
    number = {4},
    urldate = {2025-10-31},
    journal = {Superconductor Science and Technology},
    author = {Burnett, J and Faoro, L and Lindström, T},
    month = apr,
    year = {2016},
    pages = {044008},
}

@article{wang_operating_2024,
    title = {Operating semiconductor quantum processors with hopping spins},
    volume = {385},
    url = {https://www.science.org/doi/10.1126/science.ado5915},
    doi = {10.1126/science.ado5915},
    number = {6707},
    urldate = {2025-11-03},
    journal = {Science},
    author = {Wang, Chien-An and John, Valentin and Tidjani, Hanifa and Yu, Cécile X. and Ivlev, Alexander S. and Déprez, Corentin and van Riggelen-Doelman, Floor and Woods, Benjamin D. and Hendrickx, Nico W. and Lawrie, William I. L. and Stehouwer, Lucas E. A. and Oosterhout, Stefan D. and Sammak, Amir and Friesen, Mark and Scappucci, Giordano and de Snoo, Sander L. and Rimbach-Russ, Maximilian and Borsoi, Francesco and Veldhorst, Menno},
    month = jul,
    year = {2024},
    note = {Publisher: American Association for the Advancement of Science},
    pages = {447--452},
}

@article{borsoi_shared_2024,
    title = {Shared control of a 16 semiconductor quantum dot crossbar array},
    volume = {19},
    copyright = {2023 The Author(s)},
    issn = {1748-3395},
    url = {https://www.nature.com/articles/s41565-023-01491-3},
    doi = {10.1038/s41565-023-01491-3},
    
    number = {1},
    urldate = {2025-11-03},
    journal = {Nature Nanotechnology},
    author = {Borsoi, Francesco and Hendrickx, Nico W. and John, Valentin and Meyer, Marcel and Motz, Sayr and van Riggelen, Floor and Sammak, Amir and de Snoo, Sander L. and Scappucci, Giordano and Veldhorst, Menno},
    month = jan,
    year = {2024},
    note = {Publisher: Nature Publishing Group},
    pages = {21--27},
}

@article{valentini_parity-conserving_2024,
    title = {Parity-conserving {Cooper}-pair transport and ideal superconducting diode in planar germanium},
    volume = {15},
    copyright = {2024 The Author(s)},
    issn = {2041-1723},
    url = {https://www.nature.com/articles/s41467-023-44114-0},
    doi = {10.1038/s41467-023-44114-0},
    
    number = {1},
    urldate = {2025-11-03},
    journal = {Nature Communications},
    author = {Valentini, Marco and Sagi, Oliver and Baghumyan, Levon and de Gijsel, Thijs and Jung, Jason and Calcaterra, Stefano and Ballabio, Andrea and Aguilera Servin, Juan and Aggarwal, Kushagra and Janik, Marian and Adletzberger, Thomas and Seoane Souto, Rubén and Leijnse, Martin and Danon, Jeroen and Schrade, Constantin and Bakkers, Erik and Chrastina, Daniel and Isella, Giovanni and Katsaros, Georgios},
    month = jan,
    year = {2024},
    note = {Publisher: Nature Publishing Group},
    pages = {169},
}

@article{masluk_microwave_2012,
    title = {Microwave {Characterization} of {Josephson} {Junction} {Arrays}: {Implementing} a {Low} {Loss} {Superinductance}},
    volume = {109},
    shorttitle = {Microwave {Characterization} of {Josephson} {Junction} {Arrays}},
    url = {https://link.aps.org/doi/10.1103/PhysRevLett.109.137002},
    doi = {10.1103/PhysRevLett.109.137002},
    number = {13},
    urldate = {2025-11-03},
    journal = {Physical Review Letters},
    author = {Masluk, Nicholas A. and Pop, Ioan M. and Kamal, Archana and Minev, Zlatko K. and Devoret, Michel H.},
    month = sep,
    year = {2012},
    note = {Publisher: American Physical Society},
    pages = {137002},
}

@article{ambegaokar_tunneling_1963,
    title = {Tunneling {Between} {Superconductors}},
    volume = {10},
    url = {https://link.aps.org/doi/10.1103/PhysRevLett.10.486},
    doi = {10.1103/PhysRevLett.10.486},
    number = {11},
    urldate = {2025-11-03},
    journal = {Physical Review Letters},
    author = {Ambegaokar, Vinay and Baratoff, Alexis},
    month = jun,
    year = {1963},
    note = {Publisher: American Physical Society},
    pages = {486--489},
}

@article{goppl_coplanar_2008,
    title = {Coplanar waveguide resonators for circuit quantum electrodynamics},
    volume = {104},
    issn = {0021-8979},
    url = {https://doi.org/10.1063/1.3010859},
    doi = {10.1063/1.3010859},
    number = {11},
    urldate = {2025-11-03},
    journal = {Journal of Applied Physics},
    author = {Göppl, M. and Fragner, A. and Baur, M. and Bianchetti, R. and Filipp, S. and Fink, J. M. and Leek, P. J. and Puebla, G. and Steffen, L. and Wallraff, A.},
    month = dec,
    year = {2008},
    pages = {113904},
}

@article{castellanos-beltran_widely_2007,
    title = {Widely tunable parametric amplifier based on a superconducting quantum interference device array resonator},
    volume = {91},
    issn = {0003-6951},
    url = {https://doi.org/10.1063/1.2773988},
    doi = {10.1063/1.2773988},
    number = {8},
    urldate = {2025-11-03},
    journal = {Applied Physics Letters},
    author = {Castellanos-Beltran, M. A. and Lehnert, K. W.},
    month = aug,
    year = {2007},
    pages = {083509},
}

@misc{zanuz_mitigating_2024,
    title = {Mitigating {Losses} of {Superconducting} {Qubits} {Strongly} {Coupled} to {Defect} {Modes}},
    url = {http://arxiv.org/abs/2407.18746},
    
    urldate = {2024-10-28},
    publisher = {arXiv},
    author = {Zanuz, Dante Colao and Ficheux, Quentin and Michaud, Laurent and Orekhov, Alexei and Hanke, Kilian and Flasby, Alexander and Panah, Mohsen Bahrami and Norris, Graham J. and Kerschbaum, Michael and Remm, Ants and Swiadek, François and Hellings, Christoph and Lazăr, Stefania and Scarato, Colin and Lacroix, Nathan and Krinner, Sebastian and Eichler, Christopher and Wallraff, Andreas and Besse, Jean-Claude},
    month = jul,
    year = {2024},
    note = {arXiv:2407.18746 [quant-ph]},
}

@article{nsanzineza_trapping_2014,
    title = {Trapping a {Single} {Vortex} and {Reducing} {Quasiparticles} in a {Superconducting} {Resonator}},
    volume = {113},
    url = {https://link.aps.org/doi/10.1103/PhysRevLett.113.117002},
    doi = {10.1103/PhysRevLett.113.117002},
    number = {11},
    urldate = {2025-12-12},
    journal = {Physical Review Letters},
    author = {Nsanzineza, I. and Plourde, B.L.T.},
    month = sep,
    year = {2014},
    note = {Publisher: American Physical Society},
    pages = {117002},
}

@article{oehrlein_selective_1991,
    title = {Selective {Dry} {Etching} of {Germanium} with {Respect} to {Silicon} and {Vice} {Versa}},
    volume = {138},
    issn = {1945-7111},
    url = {https://iopscience.iop.org/article/10.1149/1.2085804},
    doi = {10.1149/1.2085804},
    number = {5},
    urldate = {2025-12-15},
    journal = {Journal of The Electrochemical Society},
    author = {Oehrlein, G. S. and Bestwick, T. D. and Jones, P. L. and Jaso, M. A. and Lindström, J. L.},
    month = may,
    year = {1991},
    note = {Publisher: IOP Publishing},
    pages = {1443},
}

@article{tominaga_intrinsic_2025,
    title = {Intrinsic quality factors approaching 10 million in superconducting planar resonators enabled by spiral geometry},
    volume = {12},
    issn = {2196-0763},
    url = {https://doi.org/10.1140/epjqt/s40507-025-00367-w},
    doi = {10.1140/epjqt/s40507-025-00367-w},
    number = {1},
    urldate = {2025-12-19},
    journal = {EPJ Quantum Technology},
    author = {Tominaga, Yusuke and Shirai, Shotaro and Hishida, Yuji and Terai, Hirotaka and Noguchi, Atsushi},
    month = jun,
    year = {2025},
    pages = {60},
}

\end{document}